\documentclass[a4paper,12pt]{article}

\usepackage{amssymb,latexsym,amsmath}

\usepackage{fullpage}
\usepackage{nicefrac}
\usepackage{mathrsfs}%different \cal using \mathscr
\usepackage{slashed}% for dslash
\usepackage{bbold}
\usepackage{booktabs}

\makeatletter \@addtoreset{equation}{section}
\makeatother

\begin{document}

\begin{titlepage}
	\thispagestyle{empty}
	\begin{flushright}
		\hfill{\phantom{DFPD-11/TH/XX}}
	\end{flushright}

	\vspace{35pt}

	\begin{center}
	    { \LARGE{\bf On the off-shell formulation of  $N =2$  \\[2mm] 
		supergravity with tensor multiplets}}

		\vspace{50pt}

		{Niccol\`o Cribiori and Gianguido Dall'Agata}

		\vspace{25pt}

		{
		$^1${\it  Dipartimento di Fisica e Astronomia ``Galileo Galilei''\\
		Universit\`a di Padova, Via Marzolo 8, 35131 Padova, Italy}

		\vspace{15pt}

	    $^2${\it   INFN, Sezione di Padova \\
		Via Marzolo 8, 35131 Padova, Italy}
				}

		\vspace{40pt}

		{ABSTRACT}

		\vspace{40pt}

	\end{center}

	\vspace{10pt}

		We revisit the construction of $N=2$ superconformal multiplets using rheonomic superspace techniques.
		We apply the result to the derivation of off-shell Poincar\'e supersymmetric models where a tensor multiplet couples to gravity and to an arbitrary number of vector multiplets.
		We also analyze gaugings involving the tensor field.

\end{titlepage}

\baselineskip 6 mm

\renewcommand{\arraystretch}{1.2}

%%%%%%%%%%%%%%%%%%%%%%%%%%%%%%%%%%%%%%%%%%%%%%%%%%%%%%%%%%%%%%
%%%%%%%%%%%%%%%%%%%%%%%%%%%%%%%%%%%%%%%%%%%%%%%%%%%%%%%%%%%%%%

% \pagestyle{fancy}

\section{Introduction} % (fold)
\label{sec:introduction}

Supergravity theories are the primary framework for investigating low-energy approximations of string theory and proved to be extremely important in addressing questions about the vacuum selection process, the analysis of non-perturbative phenomena and the understanding of  supersymmetry breaking mechanisms.
When string theory is compactified to four dimensions, branes and fluxes are commonly used to stabilize the vacuum and this results in generic gauged supergravity models as effective theories \cite{Samtleben:2008pe}.
The study of such models has shown that tensor fields play an important role: they are at the basis of the so-called magnetic gaugings, where, by a St\"uckelberg mechanism, they become massive by eating some combination of the gauge vectors \cite{deWit:2005ub}.

It is clear that in four dimensions one can always dualize a massless tensor field to a scalar and a massive tensor to a massive vector, but most often the natural configuration emerging from flux compactifications contains tensor fields from the very beginning, as they derive from higher rank form-fields in ten dimensions.
The duality relation needed to transform these degrees of freedom into scalars or vectors is non-perturbative \cite{Louis:2002ny}.
Moreover, the full gauge structure of the theory in the presence of tensor fields is that of a free differential algebra, rather than that of an ordinary Lie group \cite{DallAgata:2005xvr}.
Also, while the models with tensor fields and those with their scalar or vector duals are equivalent on-shell, it is not at all clear that they remain equivalent off-shell or when one considers higher order derivative corrections.
This, in turn, affects exact results that could be obtained on the computations of the black hole entropy by means of localization techniques \cite{Dabholkar:2010rm,Benini:2016rke,Hristov:2018lod}, or on the stability properties of the selected vacuum when including non-perturbative effects.
In fact, while there are natural corrections to the scalar potential that can be obtained from non-perturbative effects, it is more difficult to get similar results if the fundamental field in the theory is a tensor rather than a scalar.
For these reasons we decided to investigate once more the off-shell couplings of a tensor field to an $N=2$ extended supergravity theory, also in the presence of vector multiplets.

The couplings of tensor multiplets in generic (gauged) $N=2$ supergravity theories have been investigated before in various ways.
The general on-shell Lagrangian and supersymmetry transformations have been worked out in \cite{DallAgata:2003sjo,Sommovigo:2004vj,DAuria:2004yjt}, also in relation to flux compactifications scenarios.
Double tensor and vector-tensor multiplets were also considered in \cite{Claus:1996ze,Claus:1997fk,Claus:1998jt,Brandt:2000uw,Andrianopoli:2011zj}.
For what concerns the off-shell formulation, the main reference is \cite{deWit:2006gn}, where, using superconformal tensor calculus and building on \cite{deWit:1981fh,deWit:1982na}, the off-shell conformal coupling of tensor multiplets to supergravity has been derived and various implications of the construction have been discussed, including a derivation of the classification of 4-dimensional quaternionic-K\"ahler manifolds with two commuting isometries.
While the direct introduction of hypermultiplets forces an on-shell formulation \cite{deWit:1980lyi}, or the use of an infinite set of auxiliary fields \cite{Galperin:2001uw}, tensor multiplets can be realized off-shell also with respect to Poincar\'e supersymmetry, recovering the hypermultiplet scalar manifold geometry on-shell.

In this work we consider the off-shell formulation of the couplings of the tensor multiplet to gravity and vector multiplets also in the Poincar\'e limit.
In order to do so, we use the superconformal approach for the construction of the off-shell Lagrangian and supersymmetry rules of the $N=2$ theory, but we use a non-linear multiplet as a compensator for the conformal symmetries rather than a hypermultiplet \cite{Mohaupt:2000mj,Gruss:2007fr}.
The choice of this approach comes with a series of byproducts.
We first review the construction of the supersymmetry rules for the Weyl, chiral, vector, tensor and non-linear multiplet using the rheonomic principle.
This provides an independent derivation and a test of the expressions in the literature, which in one case need some minor correction.
We then construct the action for a model that in the Poincar\'e limit contains the gravity multiplet coupled to an arbitrary number of vector fields and a single tensor multiplet.
From this analysis we show interesting relations between the various multiplets, some of which are new to our knowledge.
We also discuss the duality relation between the tensor field in our model and the dual quaternionic manifold, pointing out possible generalizations.
Finally we construct an off-shell coupling between the vector and tensor multiplets, which corresponds to an electric gauging of the shift symmetry of the scalar dual to the tensor field.
There are two interesting aspects in this construction: it is realized off-shell, before the symmetries of the dual quaternionic-K\"ahler manifold are made explicit, and it is constructed in a frame with an existing prepotential for the vector-multiplet scalar manifold.
It would be interesting to see if one could also construct generalizations of magnetic gaugings along the lines of \cite{deWit:2005ub,deVroome:2007unr,deWit:2011gk} in this off-shell framework.

% section introduction (end)

\section{Conformal supergravity} % (fold)
\label{sec:conformal_supergravity}

\allowdisplaybreaks

\subsection{Generalities} % (fold)
\label{sub:generalities}

In this work we will follow a geometric approach to derive the supersymmetry transformations of off-shell $N=2$ conformal supergravity and its multiplets, which we will then use to construct the action of the various $N=2$ multiplets in the so-called Poincar\'e off-shell supergravity.

The superconformal algebra in four dimensions with two supercharges is an extension of the Poincar\'e superalgebra, by means of a dilatation generator $D$, special conformal transformations generators $K_a$, conformal supersymmetry generators $S^A$ and the generators of the U(2) R-symmetry group $T^A{}_B$.
The multiplets under conformal supersymmetry do not have a one-to-one correspondence with the multiplets of Poincar\'e supersymmetry.
For this reason we will first discuss separately the superconformal multiplets and only later construct supersymmetric actions and couplings for the $N=2$ Poincar\'e theory.
The way to go from one formalism to the other is to assume that some of the fields of the conformal theory are just compensators for the extra symmetries, so that we can obtain the ordinary supersymmetric action by fixing these fields to specific values.

The geometric approach we follow here has been described in detail in various instances before \cite{DAuria:2004yjt,Andrianopoli:1996cm} and for this reason we will be very short in its presentation, unless we need to explain peculiar features of the system under analysis.
The first step is to extend the fields in our theory to superfields with $N=2$ supersymmetry.
We define the curvatures of these fields by providing a realization of the superconformal algebra on them.
We then solve the Bianchi identities on these curvatures by fixing their parameterization requiring that the expansion of the curvatures along the corresponding basis in superspace is given only in terms of the physical fields.
This is the so-called rheonomic principle and implies that no additional degrees of freedom have been introduced in the theory.
The solution of the Bianchi identities allows then the derivation of the supersymmetry rules, which, from the point of view of superspace, are just Lie derivatives along the vector $\epsilon = \bar{\epsilon}^A D_A + \bar{\epsilon}_A D^A$, where $D_A$ and $D^A$ are the vectors dual to the gravitino 1-forms: $D_A(\psi^B) = D^B(\psi_A) = \delta^B_A \, {\mathbb 1}$ and ${\mathbb 1}$ is the unity in spinor space. 
Concretely:
\begin{equation}\label{Lie}
	{\cal L}_{\epsilon} A^M = \left(i_{\epsilon} d + d i_{\epsilon}\right) A^M = {\mathbb D} \epsilon^M + i_{\epsilon} F^M,
\end{equation}
where $A^M$ is the set of connections, $F^M$ the corresponding curvatures, $i_{\epsilon}$ represents the inner product with the vector $\epsilon$ and ${\mathbb D}$ is the covariant derivative with respect to the superalgebra.

Let us show explicitly this construction by applying it to the fundamental multiplet of conformal supergravity: the Weyl multiplet.

% subsection generalities (end)

\subsection{Weyl multiplet} % (fold)
\label{sub:weyl_multiplet}

The Weyl multiplet is the multiplet that contains all the fields dual to the generators of the superconformal algebra.
We have the vielbein $e^a$, dual to the translation generators $P_a$, the spin connection $\omega^{ab}$, dual to the Lorentz generators $M_{ab}$, the conformal vielbein $f^a$, dual to the generators of special conformal transformations $K_a$, the 1-form $b$, dual to the dilatation generator $D$, the gravitino 1-forms $\psi^A$, dual to the supersymmetry generators $Q_A$, the conformal gravitino 1-forms $\phi^A$, dual to the conformal supersymmetry generators $S_A$ and the R-symmetry vectors $\omega^A{}_B$ and ${\cal A}$, dual to the SU(2) $\times$ U(1) generators $T_A{}^B$ and $T_{U(1)}$.
In addition, to complete the off-shell multiplet, we have some additional auxiliary fields: a doublet of fermions $\chi^A$, an antisymmetric complex tensor $T_{ab}^{AB} = - T_{ab}^{BA}$ and a real scalar ${\cal D}$.
All these fields have a fixed charge $q_w$ under Weyl and $q_c$ under chiral U(1) transformations as summarized in the following table, where we also include the chirality of the fermion fields.
This allows us to introduce a shorthand notation for the covariant derivative, which includes the Weyl and U(1) connections $b$ and ${\cal A}$ as
\begin{equation}
	D = d - q_w \,b - i q_c \,{\cal A} + \ldots,
\end{equation}
where the dots stand for the other (Lorentz, SU(2)) connections.

\bigskip

\begin{table}
	\begin{center}
% \emph{Weyl Multiplet}
\begin{tabular}{*{12}{c}}\toprule
 & $e^a$ & $\psi^A$ & $b$ & ${\cal A}$ & $\omega^A{}_B$ & $T_{ab}^{AB}$ & $\chi^A$ & ${\cal D}$ & $f^a$ & $\phi^A$ & $\omega^{ab}$ \\\midrule
$q_w$ & -1 & -$\frac12$ & 0 & 0 & 0 & 1 & $\frac32$ & 2 & 1 & $\frac12$ & 0\\
$q_c$ & 0 & $\frac12$ & 0 & 0 & 0 & 1 & $\frac12$ & 0 & 0 & $\frac12$ & 0\\
$\gamma_5$ &&+&&&&&+&&&-&\\\bottomrule
\end{tabular}
	\end{center}
	\caption{Weyl multiplet}
	\label{weyltable}
\end{table}

The curvatures defined from the connections in Table~\ref{weyltable} provide a representation of the superconformal algebra:
\begin{align}
	\hat{T}^a &\equiv d e^a + \omega^a{}_b e^b + b\, e^a - \bar{\psi}^A \gamma^a \psi_A = D e^a - \bar{\psi}^A \gamma^a \psi_A, \\[2mm]
	\hat{K}^a &\equiv d f^a + \omega^a{}_b f^b - b\, f^a - \bar{\phi}^A \gamma^a \phi_A = D f^a  - \bar{\phi}^A \gamma^a \phi_A, \\[2mm]
	\hat{R}^{ab} &\equiv d \omega^{ab} + \omega^{a}{}_c \omega^{cb} + 4 f^{[a}e^{b]} + \bar{\psi}^A \gamma^{ab} \phi_A + \bar{\psi}_A \gamma^{ab} \phi^A \nonumber\\[2mm]
	&= R^{ab}+ 4 f^{[a}e^{b]} + \bar{\psi}^A \gamma^{ab} \phi_A + \bar{\psi}_A \gamma^{ab} \phi^A ,\\[2mm]
	\label{rhodef}
	\hat{Q}^A &\equiv d \psi^A + \frac14 \omega^{ab} \gamma_{ab} \psi^A + \frac12\, b\, \psi^A - \frac{i}{2}\, {\cal A} \,\psi^A + \omega^A{}_B \psi^B - \gamma_a \phi^A e^a \\[2mm]
	&= D \psi^A - \gamma_a \phi^A e^a, \nonumber \\[2mm]
	\hat{S}^A &\equiv d \phi^A + \frac14 \omega^{ab} \gamma_{ab} \phi^A - \frac12\, b\, \phi^A - \frac{i}{2}\, {\cal A} \phi^A + \omega^A{}_B \phi^B - \gamma_a \psi^A f^a \\[2mm]
	&= D \phi^A - \gamma_a \psi^A f^a, \nonumber \\[2mm]
	\hat{B} &\equiv db +2 e^a f_a + \bar{\psi}^A \phi_A + \bar{\psi}_A \phi^A, \\[2mm]
	\hat{\cal{F}} &\equiv d{\cal A} - i \left(\bar{\psi}^A \phi_A - \bar{\psi}_A \phi^A\right), \\[2mm]
	\hat{R}^A{}_B &\equiv d \omega^A{}_B + \omega^{A}{}_C \,\omega^C{}_B - 2 \bar{\psi}^A \phi_B - 2 \bar{\phi}^A \psi_B + \delta^A_B \left(\bar{\psi}^C \phi_C - \bar{\psi}_C \phi^C\right) \\[2mm]
	&= R^A{}_B - 2 \bar{\psi}^A \phi_B - 2 \bar{\phi}^A \psi_B + \delta^A_B \left(\bar{\psi}^C \phi_C - \bar{\psi}_C \phi^C\right). \nonumber
\end{align}
It is straightforard to check that Bianchi identities close imposing zero curvature conditions.
In order to go beyond, we need to give parameterizations of the curvatures.
We do it in a framework where we impose constraints that remove $\omega^{ab}$, $\phi^A$ and $f^a$ from the set of physical fields and make them functions of the other fields.
This is needed to obtain the match of the bosonic and fermionic degrees of freedom, which, for the Weyl multiplet are 24 each.
Actually, in order to do that, we also need to introduce the definitions of the curvatures for the  auxiliary fields:
\begin{align}
	D {\cal D} &\equiv d {\cal D} - 2 b {\cal D}, \\[2mm]
	D \chi^A &\equiv  d \chi^A + \frac14 \omega^{ab} \gamma_{ab} \chi^A - \frac32\,b\, \chi^A - \frac{i}{2} {\cal A} \chi^A + \omega^A{}_B \chi^B, \\[2mm]
	D T_{ab}^{AB} &\equiv d T_{ab}^{AB} - (b+i {\cal A}) T_{ab}^{AB} 	- 2 \omega^c{}_{[a} T_{|c|b]}^{AB}.
\end{align}
We can then solve the corresponding Bianchi identities in superspace and the result is the full parameterization of the curvatures as follows:
\begin{align}
	\hat{T}^a &= 0, \\[2mm]
	\hat{K}^a &= \frac12 \, e^b e^c K_{bc}{}^a + \left(\bar{\psi}^A \psi^B T_b{}^{ab}_{AB}+ \bar{\psi}_A \psi_B T_b{}^{ab AB}\right)\nonumber\\[2mm]
	&+\frac12\, \left(\bar{\psi}^A \gamma_c \rho_b{}^{ba}_A e^c +\bar{\psi}_A \gamma_c \rho_b{}^{ba\,A} e^c\right) -\frac12 \,e^a\left(\bar{\psi}^A \gamma^b \chi_{bA} + \bar{\psi}_A \gamma^b \chi_b^A\right)  \nonumber\\[2mm]
	&+ \frac12\, {\cal D} \, \bar{\psi}^A \gamma^a \psi_A,\\[2mm]
	\hat{R}^{ab} &=\frac12\, e^c e^d \hat{R}_{cd}{}^{ab} - \left(\bar{\psi}^A \gamma_c \rho^{ab}_A+ \bar{\psi}_A \gamma_c \rho^{ab A}\right)e^c \nonumber\\[2mm]
	&+ \left(\bar{\psi}^A \gamma_c \gamma^{ab} \chi_A + \bar{\psi}_A \gamma_c \gamma^{ab} \chi^A\right) e^c + 2 \left(\bar{\psi}^A \psi^B T^{ab}_{AB} + \bar{\psi}_A \psi_B T^{ab\,AB} \right),\\[2mm]
	\label{rhoparam}
	\hat{Q}^A &=\frac12\, e^a e^b \rho_{ab}^A -\frac12\, T_{ab}^{AB} \gamma^{ab}\gamma_c \psi_B\, e^c\\[2mm]
	\hat{S}^A &=\frac12\, e^a e^b \phi_{ab}^A -\frac18 R_{cd}{}^A{}_B \gamma^{cd}\gamma_a \psi^B e^a + \frac{i}{8} {\cal F}_{cd} \gamma^{cd} \gamma_a \psi^A e^a \nonumber\\[2mm]
	&+ \frac14 T^{c,de\,AB} \gamma_c \gamma_{de} \gamma_a \psi_B e^a + \gamma_a \psi^A \, \bar{\psi}^B \gamma^a \chi_B, \\[2mm]
 	\hat{B} &= \frac12 e^a e^b B_{ab} + \bar{\psi}^A \gamma_a \chi_A e^a + \bar{\psi}_A \gamma_a \chi^A e^a, \\[2mm]
	\hat{\cal{F}} &= \frac12 e^a e^b {\cal F}_{ab} +i\,\left( \bar{\psi}^A \gamma_a \chi_A e^a - \bar{\psi}_A \gamma_a \chi^A e^a\right), \\[2mm]
	\hat{R}^A{}_B &= \frac12 e^c e^d R_{cd}{}^A{}_B - 2 \left(\bar{\psi}^A \gamma_a \chi_B - \bar{\psi}_B \gamma_a \chi^A\right) e^a + \delta^A_B \left(\bar{\psi}^C \gamma_a \chi_C - \bar{\psi}_C \gamma_a \chi^C\right)e^a, \\[2mm]
	D {\cal D} & = e^a {\cal D}_a + \bar{\psi}^A \gamma^a \chi_{a A} + \bar{\psi}_A \gamma^a \chi_a^A, \\[2mm]
	D \chi^A &= e^a \chi_{a}^A + \frac14\, T_{b\,cd}^{AB} \gamma^{cd} \gamma^b \psi_B -\frac{i}{8} {\cal F}_{ab} \gamma^{ab} \psi^A + \frac12 {\cal D} \psi^A  \nonumber\\[2mm]
	&+ \frac12\, T_{ab}{}^{AB} \gamma^{ab} \phi_B -\frac18 R_{ab}{}^A{}_B \gamma^{ab} \psi^B,  \\[2mm]
	D T_{ab}^{AB} &= e^c T_{c\,ab}^{AB} + \bar{\psi}^{[A} \rho_{ab}^{B]}.
\end{align}
Note that imposing $T^a = 0$ tells us that $\omega^{ab}$ is a function of the vielbein, but also of $\psi^A$ and especially of $b$.
This implies that the Riemann curvature does not satisfy the standard constraints.
For instance we still have that
\begin{equation}
	\hat{R}_{a[b\,cd]} = 0,
\end{equation}
but now
\begin{equation}
	\hat{R}_{[ab}{}^c{}_{d]} = - B_{[ab} \delta^c_{d]}.
\end{equation}
As mentioned above, the closure of Bianchi identities (BI) really follows only upon imposing appropriate constraints, used to remove the extra degrees of freedom with respect to the physical ones.
They can be obtained in a fast way by inspecting the Bianchi identities at the $\psi^A \gamma^a \psi_A$ level.
From the gravitino BI we get a relation between the gravitino field strength and the spin 1/2 field in the multiplet
\begin{equation}\label{rhoconst}
	\gamma^b \rho_{ab}^A = -2 \gamma_a \chi^A, \qquad \Leftrightarrow \qquad e^a \gamma_a \rho^A = \frac{1}{3} e^a e^b e^c \gamma_{abc} \chi^A,
\end{equation}
as well as the fact that the gravitino curvature has to satisfy a self-duality condition:
\begin{equation}\label{rhoselfdual}
	\rho_{ab}^A = \frac{i}{2} \epsilon_{ab}{}^{cd} \rho_{cd}^A.
\end{equation}
From the dilatation gauge field BI  we obtain a relation between its curvature and the curvature of ${\cal A}$:
\begin{equation}\label{BFconst}
	B_{ab} = - \frac12 \epsilon_{abcd} {\cal F}^{cd} \qquad \Leftrightarrow \qquad B^+_{ab} = i \, {\cal F}^+_{ab}\,.
\end{equation}
From the conformal gravitino BI we obtain
\begin{equation}\label{phiconst1}
	\phi_{ab}^A + \frac{i}{2} \epsilon_{abcd} \phi^{cd A} = - \gamma^{ab}\gamma^c \chi_{c}^A +2  \gamma_{[a} \rho^c{}_{b]c}^A - 8 T_{ab}^{AB} \chi_B,
\end{equation}
and
\begin{equation}\label{phiconst2}
	\gamma^{ab} \phi_{ab\, A} = 2 \gamma^a \chi_{a A}.
\end{equation}
Finally, from the spin connection BI we get
\begin{equation}\label{Rconst}
	\hat{R}_{ca}{}^c{}_b + B_{ab} + i F_{ab} -2 {\cal D}\, \eta_{ab} + 8\,  T_{ac\, AB} \,T^c{}_b{}^{AB} = 0.
\end{equation}
Note that the constraints are $S$-invariant and therefore, when closing Bianchi identities, terms with $\phi^A$ cancel by themselves.
On the other hand they are not $Q$-supersymmetric invariant, but rather transform into each other by the action of supersymmetry.
For instance, (\ref{rhoselfdual}) follows from (\ref{rhoconst}), (\ref{phiconst1}) follows from (\ref{BFconst}) and (\ref{phiconst2}) follows from supersymmetry applied to the trace of (\ref{Rconst}).

The constraints (\ref{rhoconst}) and (\ref{Rconst}) are especially important, because their solution in ordinary space-time gives an expression for the conformal gravitini and for the conformal vielbein in terms of the other fields.
These can be derived by projecting the constraints from superspace to ordinary spacetime, which is obtained by replacing the curvatures in the constraints in terms of the physical fields.
The final expressions, in turn, can be read from the curvature parameterizations.
For instance, the spacetime definition of $\rho_{ab}^A$ follows from (\ref{rhoparam}) and (\ref{rhodef}) as
\begin{equation}
	\rho_{\mu\nu}^A = 2 D_{[\mu} \psi_{\nu]}^A + 2 \gamma_{[\mu} \phi^A_{\nu]} - \slashed{T}^{AB} \gamma_{[\mu} \psi_{\nu]\,B},
\end{equation}
where
\begin{equation}
	D_{\mu} \psi_{\nu}^A = \left[\partial_\mu + \frac14\, \omega_\mu{}^{ab} \gamma_{ab} - \frac12 (b_\mu + i\, {\cal A}_\mu)\right] \psi_\nu^A - \Gamma_{\mu\nu}^\rho \psi_{\rho}^A + \omega_\mu{}^A{}_B \psi_\nu^B
\end{equation}
and $\Gamma^\rho_{\mu\nu}$ is the standard Levi--Civita connection.
Plugging this into the constraint (\ref{rhoconst}), one finds 
\begin{equation}
	\phi_a^A = \frac{1}{24}\left(\gamma_a \gamma^{bc} -3 \gamma^{bc}\gamma_a\right)\left[2 \, D_b \psi_{c}^A - \slashed{T}^{AB} \gamma_{b}\psi_{c B} - \frac23 \, \gamma_{bc} \chi^A\right].
\end{equation}
Similarly, the conformal vielbein is determined by (\ref{Rconst}) once one plugs in the expression for the curvature and the other fields.
We only give here the expression for its trace, which is going to be needed in the following:
\begin{equation}
	f_a^a = - \frac{1}{12} R + \frac23 {\cal D} -\frac{1}{12} \left(\bar{\psi}_{\mu A}\gamma^{\mu\nu \rho}D_\nu \psi_{\rho}^A - 4\, \bar{\psi}_{\mu A}\psi_{\nu B} T^{\mu \nu AB} + 2\, \bar{\psi}_{\mu A} \gamma^\mu \chi^A + h.c. \right).
\end{equation}

Now that we have the full parameterizations of the Weyl multiplet it is also straightforward to deduce the transformation rules of the physical fields with respect to the conformal supergroup by direct inspection of the parameterization themselves, by means of (\ref{Lie}).
As an example we give here the transformation of the 1-form $b$ and of the conformal vielbein under special conformal transformations:
\begin{align}
	\delta_{\Lambda_k} f^a &= {\cal L}_{\Lambda_k} f^a = i_{\Lambda_k} K^a + D \Lambda_k^a = D \Lambda_k^a, \\[2mm]
	\delta_{\Lambda_k} b &= -2 e^a \Lambda_{k\,a}.
\end{align}
An analogous procedure gives all the other transformations including supersymmetry.
The only care that is needed is with respect to the fact that the constraints are not invariant under supersymmetry and therefore supersymmetry transformations of the fields removed by the constraints, namely $\omega^{ab}$, $f^a$ and $\phi^A$, change by extra terms proportional to curvature tensors.
The resulting supersymmetry transformations for the whole multiplet match the structure presented in \cite{deWit:1979dzm,deWit:1980lyi}.

% subsection weyl_multiplet (end)

\subsection{Chiral multiplet} % (fold)
\label{sub:chiral_multiplet}

We now move to the analysis of the matter multiplets.
The first multiplet we consider is the chiral multiplet, which is a generalization to the conformal extended algebra of the chiral representation one has for minimally supersymmetric models in four dimensions.
This is a general multiplet of conformal and chiral weight $q$, whose field content is summarized in Table~\ref{tab:chiral}.
There are two complex scalar fields $P$ and $C$, two spin 1/2 fields $\sigma_A$ and $\Omega_A$ and a complex self-dual tensor $t^+_{ab}$, so that 
\begin{equation}
	\left(\delta_{ab}^{cd} - \frac{i}{2} \epsilon_{ab}{}^{cd}\right) t_{cd}^+ = 0 \quad \Leftrightarrow \quad \gamma^{ab} \psi_A t_{ab}^+ = 0.
\end{equation}
We will see later that the vector multiplet is a special instance of such multiplet with $q=1$ and further constrained.

\begin{table}[ht]
\begin{center}
	\begin{tabular}{*7{c}}\toprule
	 & $P$ & $\sigma_A$ & $Z_{AB}$ & $t_{ab}^+$ & $\Omega_A$ & $C$ \\\midrule
	$q_w$  & $q$ & $q+\frac12$ & $q+1$ &$q+1$ & $q+\frac32$ & $q+2$\\
	$q_c$  & $q$ & $q-\frac12$ & $q-1$ &$q-1$ & $q-\frac32$ & $q-2$\\
	$\gamma_5$ &&+&&&+&\\\bottomrule
	\end{tabular}
\end{center}
	\caption{Chiral Multiplet}
	\label{tab:chiral}
\end{table}

The curvature definitions for the chiral multiplet are
\begin{align}
	DP &= dP -q \left(b + i \,{\cal A}\right) P, \\[2mm]
	D \sigma_A &= d \sigma_A + \frac14 \, \omega^{ab} \gamma_{ab} \sigma_A - \omega^B{}_A \sigma_B - \left(q+\frac12\right) b \, \sigma_A - i \left(q-\frac12\right){\cal A} \sigma_A, \\[2mm]
	D Z_{AB} &= \left[d - (q+1)b - i (q-1) \,{\cal A}\right] Z_{AB} - 2 \omega^C{}_{(A} Z_{B)C}, \\[2mm]
	D t_{ab}{}^+ &= \left[d - (q+1)b - i (q-1) \,{\cal A}\right] t_{ab}{}^+ - 2\,\omega^c{}_{[a} t_{|c|b]}{}^+, \\[2mm]
	D \Omega_A &= \left[d +\frac14 \, \omega^{ab} \gamma_{ab} - \left(q+\frac32\right)b - i \left(q-\frac32\right) {\cal A}\right] \Omega_A- \omega^B{}_A \Omega_B, \\[2mm]
	DC &= \left[d - (q+2)b - i (q-2)\, {\cal A}\right] C,
\end{align}
while the parameterizations that satisfy the BI following from the definitions above are
\begin{align}
	DP &= e^a P_a+i\, \bar{\psi}^A \sigma_A, \\[2mm]
	D \sigma_A &=  e^a \sigma_{aA} - i\,\gamma^a \psi_A \, P_a + \frac12 \epsilon_{AB} \, \slashed{t}^+ \psi^B + Z_{AB} \psi^B + 2 \,i q\, P\, \phi_A\,, \\[2mm]
	D Z_{AB} &= e^a Z_{a\, AB} + \bar{\psi}_{(A} \gamma^a \sigma_{a B)} + \bar{\psi}^{C} \Omega_{(A} \epsilon_{B) C} + 2 (q-1) \bar{\phi}_{(A} \sigma_{B)}, \\[2mm]
	D t_{ab}{}^+ &=e^c t_{c\,ab} + \frac14\, \epsilon^{AB}\, \bar{\psi}_A \gamma^c \gamma_{ab} \sigma_{cB} -\frac14 \bar{\psi}^A \gamma_{ab} \Omega_A + \frac12 (q+1) \, \epsilon^{AB} \bar{\phi}_A \gamma_{ab} \sigma_B, \\[2mm]
	D \Omega_A &= e^a \Omega_{aA} +\frac12 \, \gamma^{ab}\gamma^c \psi_A \, t_{c\,ab} +  \gamma^a \psi_C\, Z_{a AB} \epsilon^{BC} + C\, \epsilon_{AB} \psi^B + \nonumber\\[2mm]
	&+i\,  \gamma^a \gamma^{bc} \psi_C\, P_a T_{bc\, AB} \epsilon^{BC} + i \,q\, P\, \gamma^a \gamma^{bc} \psi_C\,  \epsilon^{BC} \, T_{a\,bc AB} \\[2mm]
	&-4 \,\bar{\chi}_{[A} \gamma^a \sigma_{B]}\, \epsilon^{BC} \, \gamma_a \psi_C - 2 (q+1) Z_{AB} \epsilon^{BC} \phi_C - (q-1) \gamma^{ab}\phi_A \, t_{ab}^+, \nonumber \\[2mm]
	 DC &= e^a C_a - \epsilon^{AB} \bar{\psi}_A \gamma^a \Omega_{aB} -8 \bar{\psi}_A \chi_B \, \epsilon^{AC} \epsilon^{BD} Z_{CD} \nonumber \\[2mm]
	 &+ \epsilon^{AB} \epsilon^{CD} (q-1) \bar{\psi}_A \gamma^{ab} \gamma^c \sigma_D \, T_{c\,ab\,BC} + \epsilon^{AB} \epsilon^{CD}  \bar{\psi}_A \gamma^{ab}\gamma^{c} \sigma_{cD} T_{ab\, BC} \\[2mm]
	 &- 2 \, q\, \epsilon^{AB} \bar{\phi}_A \Omega_B. \nonumber
\end{align}
Also in this case the supersymmetry transformation rules we deduce from this construction match the ones in \cite{deWit:1979dzm,deRoo:1980mm,deWit:1980lyi}, modulo differences that are related to different conventions.

% subsection chiral_multiplet (end)

\subsection{Vector multiplet} % (fold)
\label{sub:vector_multiplet}

As mentioned in the previous subsection, the vector multiplet is a special instance of chiral multiplet.
The matter content is restricted to a complex scalar field $L$, two fermions $\lambda_A$, a vector field $W_\mu$ and a triplet of auxiliary fields $Y_{AB} = Y_{BA}$, satisfying
\begin{equation}
	Y_{AB} = \epsilon_{AC} \epsilon_{BD} Y^{CD}.
\end{equation}
The charges in our conventions are presented in Table~\ref{tab:vector}.

\begin{table}[ht]
	\begin{center}
	\begin{tabular}{*5{c}}\toprule
	 & $W_\mu$ & $Y_{AB}$ & $\lambda_A$ & $L$ \\\midrule
	$q_w$  & 0 & 2 & $\frac32$ & 1\\
	$q_c$  & 0 & 0 & $\frac12$ & 1\\
	$\gamma_5$ &&&+&\\\bottomrule
	\end{tabular}
	\end{center}
	\caption{Vector Multiplet}
	\label{tab:vector}
\end{table}

Since we are going to consider an arbitrary number $n_V+1$ of such multiplets, one of which is going to play the role of compensator, in the following we will add an index $\Lambda,\Sigma,\ldots = 0,1, \ldots, n_V$ to the fields.
The curvature definitions for such multiplets are the following
\begin{align}
	F^\Lambda &\equiv dW^\Lambda -\frac12  L^\Lambda \bar{\psi}_A \psi_B\epsilon^{AB}  -\frac12 \bar{L}^\Lambda \bar{\psi}^A \psi^B \epsilon_{AB},\\[2mm]
	DY_{AB}^\Lambda &\equiv dY_{AB}^\Lambda -2b Y_{AB}^\Lambda - 2 \omega^C{}_{(A}Y^{\Lambda}_{|C|B)},  \\[2mm]
	D \lambda_A^\Lambda &\equiv d \lambda_A^\Lambda +\frac14\, \omega^{ab} \gamma_{ab} \lambda_A^\Lambda -\frac32 b \lambda_A^\Lambda -\frac{i}{2} {\cal A} \lambda_A^\Lambda- \omega^B{}_A \lambda_B^\Lambda, \\[2mm]
	DL^\Lambda &\equiv dL^\Lambda -(b+i\,{\cal A}) L^\Lambda,
\end{align}
while their rheonomic parameterization, solving the corresponding Bianchi identities is
\begin{align}
	F^\Lambda &= \frac{1}{2} e^a e^b \left({\cal F}_{ab}^\Lambda +L^\Lambda T_{abAB} \epsilon^{AB}+  \bar{L}^{\Lambda} T_{ab}^{AB}\epsilon_{AB}\right)\nonumber\\[2mm]
	&- \frac12 e^a \bar{\psi}_A \gamma_a \lambda_B^\Lambda \epsilon^{AB}-\frac12 e^a \bar{\psi}^A \gamma_a \lambda^{B\Lambda}  \epsilon_{AB},\\[2mm]
	DY_{AB}^\Lambda &= e^a Y_{a AB}^\Lambda + \bar{\psi}_{(A}\gamma^a \lambda^{\Lambda}_{a B)} + \epsilon_{(A|C|}\epsilon_{B)D} \bar{\psi}^C \gamma^a \lambda_a^{\Lambda D},\\[2mm]
	\label{Dlambda}
	D \lambda_A^\Lambda &= e^a \lambda_{a A}^\Lambda+ \gamma^a \psi_A \,L_a^\Lambda + \frac12\, \epsilon_{AB} \slashed{\cal F}^\Lambda \psi^B + Y_{AB}^\Lambda \psi^B - 2 L^\Lambda \phi_A,\\[2mm]
	\label{DX}
	DL^\Lambda &= e^a L_a^\Lambda + \bar{\psi}^A \lambda_A^\Lambda.
\end{align}
The construction of the action for this multiplet requires also the following parameterization 
\begin{align}
		DL_a^\Lambda &= e^b S_{ab}^\Lambda - 2 f_a L^\Lambda + \frac{L^\Lambda}{2}\left(B_{ab} + {\cal F}_{ab}\right) e^b
	 + \frac12 \, \bar{\lambda}^\Lambda_A \rho_{ab}^A \, e^b + \bar{\psi}^A \lambda_{aA}^\Lambda- \bar{\lambda}_A^\Lambda \gamma_a \phi^A  \\[2mm]
	 	&-2L^\Lambda\, \bar{\psi}_A \gamma_a \chi^A + \frac12 \bar{\lambda}_A^\Lambda \slashed{T}^{AB}\gamma_a \psi_B, \nonumber
\end{align}
where $S_{ab}^\Lambda = S_{ba}^\Lambda$.
Also in this case, up to differences due to conventions, the result agrees with \cite{deWit:1979dzm,deRoo:1980mm,deWit:1980lyi}.

% subsection vector_multiplet (end)

\subsection{Tensor multiplet} % (fold)
\label{sub:tensor_multiplet}

The multiplet we are more interested in this work is the tensor multiplet, presented in Table~\ref{tab:tensor}.
This contains a 2-form $B$, a triplet of scalars $L^{AB}=L^{BA}$, satisfying
\begin{equation}
	L_{AB} = \epsilon_{AC} \epsilon_{BD} L^{CD},
\end{equation}
a complex auxiliary field $G$ and a doublet of spin 1/2 fields $\zeta^A$.

\begin{table}[ht]
	\begin{center}
	\begin{tabular}{*5{c}}\toprule
	 & $B$ & $L^{AB}$ & $\zeta^A$ & $G$ \\\midrule
	$q_w$  & 0 & 2 & $\frac52$ & 3\\
	$q_c$  & 0 & 0 & $-\frac12$ & -1\\
	$\gamma_5$&&&+&\\\bottomrule
	\end{tabular}
	\end{center}
	\caption{Tensor Multiplet}
	\label{tab:tensor}
\end{table}

The curvature definitions are
\begin{align}
	H &\equiv dB+ i L_{AB} \epsilon^{BC} \bar{\psi}^A \gamma_a \psi_C e^a,\\[2mm]
	DL^{AB} &\equiv dL^{AB} -2b L^{AB} +2 \omega^{(A}{}_C L^{B)C},  \\[2mm]
	D \zeta^A &\equiv d \zeta^A +\frac14\, \omega^{ab} \gamma_{ab} \zeta^A -\frac52 b \,\zeta^A +\frac{i}{2}\, {\cal A} \zeta^A+ \omega^A{}_B \zeta^B, \\[2mm]
	DG &\equiv dG -3b\,G + i \,{\cal A} G
\end{align}
and the rheonomic parameterization is
\begin{align}
	H &= \frac{1}{6} e^a e^b e^c H_{abc} +\frac{i}{4}  \left(\bar{\psi}^A \gamma_{ab} \zeta^B \epsilon_{AB} - \bar{\psi}_A \gamma_{ab} \zeta_B \epsilon^{AB}\right)e^a e^b,\\[2mm]
	DL^{AB} &= e^a L_a^{AB} + \bar{\psi}^{(A}\zeta^{B)} + \epsilon^{(A|C|}\epsilon^{B)D} \bar{\psi}_C \zeta_D, \\[2mm]
	D \zeta^A &= e^a \zeta_a^A - 4 L^{AC} \phi_C +\gamma^a \psi_B L_a^{AB} + \epsilon^{AB} \tilde{H}^a \gamma_a\psi_B + G \psi^A,\\[2mm]
	DG &= e^a G_a + 2 \bar{\phi}_A \zeta^A + \bar{\psi}_A \gamma^a \zeta_a^A -8 \,L^{AB} \bar{\psi}_A \chi_B -  \bar{\psi}_A \gamma^{cd} \zeta_B \, T_{cd CD} \epsilon^{AC} \epsilon^{BD},
\end{align}
where we introduced the notation
\begin{equation}
	\tilde{H}^a = \frac16 \epsilon^{abcd} H_{bcd}.
\end{equation}
Once more the result agrees with \cite{deWit:1979dzm,deWit:1980lyi,deWit:1981fh,deWit:1982na,deWit:2006gn}, up to differences in the conventions.

% subsection tensor_multiplet (end)

\subsection{Non-linear multiplet} % (fold)
\label{sub:non_linear_multiplet}

The last multiplet we consider is the non-linear multiplet.
This is a peculiar multiplet, whose fields have non-linear transformations under supersymmetry even off-shell.
Its main purpose in this work is to provide the degrees of freedom needed as compensators of the SU(2) transformations of the superconformal extended group.

The field content is summarized in Table~\ref{tab:nonlinear}.
There is a triplet of scalar fields, arranged in a 2 by 2 matrix, with a non-trivial action of the superconformal SU(2) subgroup on one side (acting on the $A,B,C,\ldots$ indices) and of a flavour SU(2) on the other (acting on the $i,j,k,\ldots$ indices).
For this reason it satisfies the following relations:
\begin{align}
	\Phi^A_i \Phi_A^j  = \delta^{j}_i, \quad
	\Phi^A_i \Phi_B^i  = \delta^{A}_B,  \quad
	\Phi^A_i = \epsilon^{AB} \epsilon_{ij} \Phi_B^j, \quad
	\Phi^A_i \Phi_{aA}^i =0.
\end{align}
We then have a doublet of fermion fields $\tau^A$, a complex field $M^{AB} = - M^{BA}$ and a real vector field $V^a$.
Since the bosonic and fermionic degrees of freedom do not match, we have to impose a constraint on the vector field to reduce its degrees of freedom by one.

\begin{table}[ht]
	\begin{center}
	\begin{tabular}{*5{c}}\toprule
	 & $\Phi_A^i$ & $M^{AB}$ & $\tau^A$ & $V_a$ \\\midrule
	$q_w$  & 0 & 1 & $\frac12$ & 1\\
	$q_c$  & 0 & 1 & $\frac12$ & 0\\
	$\gamma_5$&&&-&\\\bottomrule
	\end{tabular}
	\end{center}
	\caption{Non-linear Multiplet}
	\label{tab:nonlinear}
\end{table}

The curvature definitions are
\begin{align}
	D \Phi^A_i &\equiv d \Phi^A_i + \omega^A{}_B \Phi^B_i, \\[2mm]
	DM^{AB} &\equiv d M^{AB} - (b+i\,{\cal A})M^{AB},  \\[2mm]
	D \tau^A &\equiv d \tau^A + \frac14 \omega^{ab} \gamma_{ab} \tau^A - \frac12 (b+i\,{\cal A}) \tau^A + \omega^A{}_B \tau^B, \\[2mm]
	D V_a &\equiv d V_a - \omega^b{}_a V_b - b V_a.
\end{align}
We can then solve the Bianchi identities following from such curvatures, which leads to the parameterization
\begin{align}
	D \Phi^A_i &= e^a \Phi_{a\,i}^A+ (2 \bar{\psi}^A \tau_B - \delta^A_B \bar{\psi}^C \tau_C -2 \bar{\psi}_B \tau^A + \delta^A_B \bar{\psi}_C \tau^C)\Phi^B_i,\\[2mm]
	DM^{AB} &= e^a M_a^{AB} + 4 \, \bar{\psi}^{[A} \chi^{B]} - \bar{\psi}^{C}\gamma^{ab}\tau_C T_{ab}^{AB} -4\, \bar{\psi}^{[A}\gamma^a \tau^{B]} V_a - 2\, \bar{\psi}^{C}\tau_C M^{AB}\nonumber\\[2mm]
		&-2 \bar{\psi}^{[A}\gamma^a \tau_a^{B]} -2  \bar{\psi}^{[A}\Phi^{B]}_i \gamma^a \Phi_{a\,C}^i \tau^C, \\[2mm]
	D \tau^A &= e^a \tau_a^A + \gamma^a \psi^A V_a + M^{AB} \psi_B + \frac12 \Phi^A_i \gamma^a \psi^B \Phi_{aB}^i  - \phi^A \nonumber\\[2mm]
	&-2 \,\tau^A \left(\bar{\psi}_D\tau^D + \bar{\psi}^D \tau_D\right) + \gamma^a\psi^A\, \bar{\tau}^B \gamma_a \tau_B + \frac12 \, \gamma_{ab} \psi_B \, \bar{\tau}^B \gamma^{ab} \tau^A,\\[2mm]
	D V_a &= e^b V_{ba}+\left[ \frac12 \bar{\psi^A} \gamma^a \chi_A +\frac14 \bar{\psi}^A \gamma_a \gamma_{bc} \tau^B T^{bc}_{AB} - \bar{\psi}^A \gamma_a \gamma^b \tau_A V_b + \bar{\psi}^A \gamma_a \tau^B M_{AB}\right.\nonumber \\[2mm]
	&-\left. \frac12 \bar{\psi}^A \gamma_{ab} \tau^{b}_A -\frac12  \Phi_i^A\, \bar{\psi}_A \gamma_a \gamma^b \tau^B \Phi_{bB}^i -\frac12 \bar{\tau}_A \gamma_a \phi^A + h.c. \right] - f_a.
 \end{align}

In order to check the Bianchi identities one has to impose the constraint
\begin{align}\label{nonlinearconst}
	V_a^a &= {\cal D} - 2 V_a V^a - M_{AB}M^{AB} + \frac14 \Phi_{ai}^A \Phi^{ai}_A + \bar{\tau}^A \gamma^a \tau_{a A} + \bar{\tau}_A \gamma^a \tau_a^A - \bar{\tau}_A \slashed{T}^{AB} \tau_B  \nonumber\\[2mm]
	&-  \bar{\tau}^A \slashed{T}_{AB} \tau^B-2 \bar{\tau}^A \chi_A -2 \bar{\tau}_A \chi^A +2 \bar{\tau}^A \gamma^a \tau_B \, \Phi^B_{ai} \Phi_A^i.
\end{align}
This constraint is invariant under special conformal transformations, supersymmetry and conformal supersymmetry.
The last term is especially needed to prove invariance under conformal supersymmetry and supersymmetry.
Our result crucially differs from the one in \cite{deWit:1980lyi} because of the presence of this additional term in the constraint.

% subsection non_linear_multiplet (end)

% section conformal_supergravity (end)

\section{Actions} % (fold)
\label{sec:actions}

Now that we solved all the Bianchi identities of the various multiplets, we can construct Lagrangians for the models we are interested in.
These are models which produce the couplings of supergravity to an arbitrary number of vector multiplets and one tensor multiplet off-shell, but without preserving conformal invariance.
For this reason we proceed in steps.
First we construct the superconformal couplings of an arbitrary number of vector multiplets and of a tensor multiplet.
Then we introduce the necessary compensators, which we are going to fix, in order to break conformal invariance and obtain a realization of the super-Poincar\'e group on the action.

While the actions we produce are written in superspace, the ordinary spacetime sections are obtained by identifying all the components in the rheonomic expansion with the appropriate spacetime fields, projecting the parameterizations of the curvatures on the $\theta = 0 = d \theta$ surface.
This is needed because all our forms are expressed in superspace, which means that one should in principle expand the supervielbein $E^I = (e^a, \psi^A, \psi_A)$ and all other fields in superspace, i.e.~$e^a(x, \theta) = e^a_\mu(x, \theta) dx^\mu + \bar{e}^a_A(x,\theta) d \theta^A + \bar{e}^{aA}(x, \theta) d \theta_A$ for the vielbein and analogous expressions for the other superforms.
Invariance under the various symmetries is then straightforward in the superspace approach and can be easily verified by taking a (covariant) differential of the integrated 4-form and checking that it vanishes.
For supersymmetry this amounts to actually check that each different projection on the gravitini is vanishing separately \cite{Andrianopoli:1996cm}.
Analogously, for the conformal supersymmetry, this amounts to check that each different projection on the conformal gravitini is vanishing.

The multiplets of conformal supergravity are not in one to one correspondence with the multiplets of ordinary Poincar\'e supergravity.
For instance, the Weyl multiplet, which contains the graviton and the gravitini, does not contain the graviphoton, which is a fundamental field of the $N=2$ supergravity multiplet.
This means that in order to construct (off-shell) Poincar\'e supergravity using superconformal calculus, we need to at least couple a vector multiplet to the Weyl multiplet, so that we can introduce the degrees of freedom corresponding to the graviphoton.
This is indeed what is called the minimal field representation and it is the common part to each of the different possible realizations of Poincar\'e supergravity from superconformal symmetry.
Actually one can break superconformal invariance, fixing the compensator fields associated to the symmetries of the superconformal group that are not present in the Poincar\'e group, at least in three different ways \cite{deWit:1982na}.
These are obtained by introducing a compensating hypermultiplet, a compensating tensor multiplet or a compensating non-linear multiplet .
While all three choices are legitimate, the fixing of the dilatational symmetry in the first two cases imposes conditions that are not supersymmetric invariant and that force the use of the equations of motion, resulting in on-shell Poincar\'e models \cite{deWit:1982na,Mohaupt:2000mj}.
This is a consequence of the fact that in the Lagrangian there are terms linear in the ${\cal D}$ auxiliary field.
The non-linear multiplet on the other hand can be used in a way that removes such linear terms and therefore one can remain off-shell also after the fixing of the compensator fields.
This is the reason why in the following we will use this multiplet to construct the full action.

\subsection{Vector multiplets action} % (fold)
\label{sub:vector_multiplets_action}

As explained above, if we want to couple one or more physical vector multiplets to supergravity in the Poincar\'e action, we need at least two vector multiplets in the conformal approach, one of the two containing the graviphoton of $N=2$ Poincar\'e supergravity.
For the sake of generality we then decided to give here the action for a general number of vector multiplets, one of which will play the role of compensator.

Vector multiplets are restricted chiral multiplets and an interesting property that follows from this is that an arbitrary holomorphic function of their lowest component field $L^\Lambda$, homogeneous of degree 2 in $L^\Lambda$, can be identified with the lowest component of a chiral multiplet with Weyl and chiral charge $q_w=q_c=2$ \cite{deWit:1980lyi}.
If we identify the lowest component of the chiral field with such holomorphic function as follows
\begin{equation}
	P = P(L),
\end{equation}
and $P_{\Lambda}$, $P_{\Lambda \Sigma}$ and $P_{\Lambda \Sigma \Gamma}$ with the first, second and third derivative of $P$ with respect to $L^\Lambda$, we obtain that the closure of the vector fields BI imposes the closure of the BI of the corresponding chiral multiplet.
This happens if we identify the various components of the chiral multiplet in terms of the vector multiplet fields as follows:
\begin{align}\label{sigmaA}
	\sigma_A &= -i P_{\Lambda} \lambda^\Lambda_A, \\[2mm]
	{t_{ab}}^+ &=-i\,P_{\Lambda} {\cal F}_{ab}^{+\Lambda} + \frac{i}{8} P_{\Lambda \Sigma}\, \epsilon^{CD} \, \bar{\lambda}^\Sigma_C \gamma^{ab} \lambda_D^\Lambda , \\[2mm]
	Z_{AB} &= -i P_{\Lambda} Y^\Lambda_{AB} +\frac{i}{2} P_{\Lambda \Sigma}\, \bar{\lambda}_A^\Lambda \lambda_B^\Sigma , \\[2mm]
	\Omega_A &= -i\, P_{\Lambda \Sigma} \, Y_{AB}^\Lambda \,\lambda_C^\Sigma\, \epsilon^{BC} -i\,P_{\Lambda} \gamma^a \lambda_a^{\Lambda B} \epsilon_{AB} - \frac{i}{2}\, P_{\Lambda \Sigma}\, \slashed{\cal F}^{+\Lambda} \lambda_A^\Sigma + \frac{i}{3}\, P_{\Lambda \Sigma \Gamma} \lambda_C^\Gamma \, \bar{\lambda}^\Lambda_A \lambda^\Sigma_B\, \epsilon^{BC}, \\[2mm]
	C &= -\frac{i}{2}\,P_{\Lambda \Sigma} \, Y_{AB}^\Lambda Y^{AB\, \Sigma} + i\, P_{\Lambda \Sigma} {\cal F}^{+\Lambda}_{ab} {\cal F}^{ab + \Sigma} - i\, P_{\Lambda} \bar{S}^{\Lambda a}_a -2i\, P_{\Lambda} {\cal F}^{-ab \Lambda} T_{ab \, CD} \epsilon^{CD} \nonumber\\[2mm]
	&+i\, P_{\Lambda \Sigma}\, \bar \lambda_A^\Lambda \gamma^a \lambda_a^{A\,\Sigma} +\frac{i}{2} P_{\Lambda \Sigma \Gamma}\, Y^{AB\, \Lambda} \bar \lambda_A^\Gamma \lambda_B^\Sigma + 4i\, P_\Lambda \bar\lambda^{A\,\Lambda}\chi_A \label{C}\\[2mm]
	& -\frac{i}{4} P_{\Lambda\Sigma\Gamma}\, {\cal F}_{ab}^{+\Lambda} \,(\bar \lambda_A^\Gamma \gamma^{ab}\lambda_B^\Sigma\, \epsilon^{AB})+ \frac{i}{96} P_{\Lambda \Sigma \Gamma\Delta}(\bar \lambda_A^\Lambda \gamma^{ab}\lambda_B^\Sigma\, \epsilon^{AB})(\bar \lambda_C^\Gamma \gamma_{ab}\lambda_D^\Delta\, \epsilon^{CD}). \nonumber
\end{align}
Note that the homogeneity of $P(L)$ with respect to $P^\Lambda$ implies that
\begin{equation}
	2 P(L) = P_{\Lambda} L^\Lambda, \quad P_{\Lambda} = P_{\Lambda \Sigma} L^\Sigma, \quad P_{\Lambda \Sigma \Gamma} L^\Gamma = 0, \quad P_{\Lambda \Sigma \Gamma \Delta} L^\Delta = - P_{\Lambda \Sigma \Gamma}.
\end{equation}
The superspace invariant action is then 
\begin{align}
	S_{vector} &= \int \left[\frac{1}{4!} e^a e^b e^c e^d \epsilon_{abcd} \left(C - 2i P \, T_{ef\,AB} \epsilon^{AB} T^{ef}_{CD} \epsilon^{CD}  \right) \right. \nonumber\\[2mm]
	&+ \frac{i}{3!}\,e^a e^b e^c \left( \epsilon^{AB} \bar{\psi}_A \gamma_{abc} \Omega_B - \bar{\psi}_A \slashed{T}_{BC} \gamma_{abc} \sigma_D \epsilon^{AB} \epsilon^{CD} \right) \\[2mm]
	&+ \frac{i}{2}\, e^a e^b\,\bar{\psi}_A \gamma_{ab} \psi_{B} Z_{CD} \epsilon^{AC} \epsilon^{BD} + i\,\epsilon^{AB}\bar{\psi}_A \psi_B \,e^a e^b\, \left(t_{ab}^{+} -2 i\, P\, T_{ab CD} \epsilon^{CD}\right) \nonumber \\[2mm]
	&\left.+i\, \epsilon^{AB} \epsilon^{CD} \bar{\psi}_A \psi_B \bar{\psi}_C \gamma_d \sigma_D e^d +\, \epsilon^{AB} \epsilon^{CD} \bar{\psi}_A \psi_B \bar{\psi}_C \psi_D P\right] + h.c., \nonumber 
\end{align}
where one replaces the expression for the various components as in (\ref{sigmaA})--(\ref{C}).

The spacetime action follows by projecting this expression, also using the constraints for the conformal vielbein and gravitini.
We give here the bosonic part of the resulting action, written using the prepotential
\begin{equation}
	\chi_V = i \left(P_{\Lambda}\bar{L}^\Lambda - \bar{P}_{\Lambda}L^\Lambda\right) = - 2\,(\Im P)_{\Lambda \Sigma} L^\Lambda \bar{L}^\Sigma,
\end{equation}
where $(\Im P)$ denotes the (positive definite) imaginary part of the matrix $P_{\Lambda \Sigma}$:
\begin{align}\label{vectoraction}
	S_{vector} = &\int d^4 x\, \sqrt{-g}\, \left[\frac{\chi_V}{6}\, R - \frac43 \chi_V {\cal D} -2\,(\Im P)_{\Lambda \Sigma} D^a L^\Lambda D_a \bar{L}^\Sigma + i\, P_{\Lambda \Sigma} {\cal F}^{+\Lambda}_{ab} {\cal F}^{+\Sigma\, ab}\right. \nonumber \\[2mm]
	&\left.- i\, \bar{P}_{\Lambda \Sigma} {\cal F}^{-\Lambda}_{ab} {\cal F}^{-\Sigma\, ab}-2 i\, P_{\Lambda} {\cal F}^{\Lambda}_{ab} T^{ab}_{AB}\epsilon^{AB}+2 i\, \bar{P}_{\Lambda} {\cal F}^{\Lambda}_{ab} T^{ab\,AB}\epsilon_{AB}\right.\\[3mm]
	&\left.+2i\, \bar{P} (T_{ab}^{AB} \epsilon_{AB})(T^{abAB} \epsilon_{AB})-2i \,P (T_{ab AB} \epsilon^{AB})(T^{ab}_{AB} \epsilon^{AB}) + (\Im P)_{\Lambda \Sigma}Y^{\Lambda}_{AB}Y^{\Sigma AB}\right]. \nonumber
\end{align}
Note that in all these formulae
\begin{equation}
	{\cal F}_{ab}^\Lambda = F_{ab}^\Lambda - L^\Lambda\, T_{ab AB}\epsilon^{AB} - \bar{L}^\Lambda \, T_{ab}^{AB}\epsilon_{AB},
\end{equation}
at the bosonic level, and $F_{\mu\nu}^\Lambda = 2 \partial_{[\mu} W_{\nu]}^\Lambda$.

While the structure resembles that of special geometry for the scalar fields of the vector multiplets and their couplings, we should emphasize that only after performing the fixing of the compensator fields and going on-shell one really recovers all the correct structures predicted by $N=2$ supersymmetry.
We should also note that in this approach the action is constructed by starting from a holomorphic function $P(L)$, which is related to the prepotential of special geometry.
It is therefore clear that one cannot produce straightforwardly Lagrangians in symplectic frames where the prepotential does not exist.
This is crucial, for instance, whenever one wants to obtain partial supersymmetry breaking \cite{Ferrara:1995xi} in gauged models and therefore this has to be obtained in a different way here.
One solution is to assume that the gaugings that are electric in the frame where a prepotential does not exist can be rotated to magnetic gaugings in a different frame, where the prepotential exists instead.
This is another compelling reason to introduce tensor multiplets from the beginning and couple them to supergravity and vector multiplets, so that we can reproduce such models.

Note also in (\ref{vectoraction}) the presence of a linear term in the auxiliary ${\cal D}$ field and the fact that the Einstein--Hilbert term comes multiplied by a function of the vector fields $\chi_V$, which has to be fixed to a specific value to obtain the Poincar\'e limit, introducing a scale that breaks conformal invariance.

% subsection vector_multiplets_action (end)

\subsection{Tensor multiplet action} % (fold)
\label{sub:tensor_multiplet_action}

The construction of the action for the tensor multiplet uses an interesting trick noted in \cite{deWit:1982na}.
Products of tensor multiplets can be recast in the form of a reduced chiral multiplet by means of a scalar function of their scalar fields $\chi_T = \chi_T(L)$, introducing an appropriate fermionic completion for the combination
\begin{equation}\label{XT0}
	L_T = \frac{1}{\chi_T} \bar{G} + \ldots,
\end{equation}
which has the correct charges and dimension to represent the lowest component of a vector multiplet.
The complex scalar field $L_T$ has indeed $q_w=q_c=1$, like the scalar of the vector multiplet, and its Bianchi identity becomes identical to (\ref{DX}), for a $\lambda_T^A$ that also satisfies (\ref{Dlambda}), with $Y_{AB}^T$ and ${\cal F}_{ab}^T$ also satisfying their respective Bianchi identities.
For a single tensor, the generating potential $\chi_T$ with $q_w = 2$ and $q_c = 0$, homogeneous in the tensor scalar fields and SU(2) invariant, is uniquely singled out to
\begin{equation}\label{singletensorSU2}
	\chi_T = \sqrt{2\,L_{AB}L^{AB}}.
\end{equation}
Consequently, the lowest component of the corresponding reduced chiral multiplet is
\begin{equation}\label{XT}
	L_T = \frac{1}{\chi_T} \bar G + \frac{1}{\chi_T^3} \, \bar{\zeta}_A \zeta_B\, L^{AB},
\end{equation}
while the other components of the multiplet are
\begin{align}
	\lambda_A^T=&\frac{1}{\chi_T}\left(\gamma^a\zeta_{aA}-\frac12\gamma^{cd}T_{cd}^{CD}\epsilon_{CD}\epsilon_{AB}-8L_{AB}\chi^B\right)\nonumber\\[2mm]
	&+\frac{1}{\chi_T^3}\left(-2L_{AB}\overline G\zeta^B - 2 \gamma^a\zeta_C L_{aAB}L^{BC}+2\epsilon_{AB}\tilde{\slashed{h}}\zeta_CL^{BC}+\bar\zeta_A\zeta_B\,\zeta^B\right)\\[2mm]
	&-\frac{6}{\chi_T^5}L_{AB}\zeta^B\, \bar\zeta_M\zeta_N L^{MN},\nonumber\\[2mm]
Y_{AB}^T&=\frac{1}{\chi_T}\left(D^aL_{aAB}-4 \mathcal{D}L_{AB}+4\bar\chi_{(A}\zeta_{B)}\right)\nonumber\\[2mm]
&+\frac{1}{\chi_T^3}\bigg[-2G\bar G L_{AB}-2\left(L_{aAD}-\epsilon_{AD}\tilde{h}_a\right) L^{DC}\left(L^a_{BC}-\epsilon_{BC}\tilde{h}^a\right)\nonumber\\[2mm]
&+G\, \bar\zeta_{A}\zeta_{B} + \bar G\, \epsilon_{AC}\epsilon_{BD}\bar\zeta^C\zeta^D+2L_{C(A}\bar\zeta^C\gamma^a \zeta_{aB)}+2L_{C(A}\bar\zeta^C_a \gamma^a\zeta_{B)}\nonumber\\[2mm]
&-2L_{AB}\bar\zeta_a^C\gamma^a\zeta_C+L_{aAB}\bar\zeta^C\gamma^a\zeta_C+L_{aC(A}\bar\zeta^C\gamma^a\zeta_{B)}+\tilde h_a \epsilon_{C(A}\bar\zeta^C\gamma^a\zeta_{B)}\nonumber\\[2mm]
&-\frac12 L_{AB}\left(T_{ab}^{CD}\epsilon_{CD}\bar\zeta^E\gamma^{ab}\zeta^F\epsilon_{EF}+T_{ab\, CD}\epsilon^{CD}\bar\zeta_E\gamma^{ab}\zeta_F\epsilon^{EF}\right)\nonumber\\[2mm]
&-16 L_{(A|C|}L_{B)D}\bar\chi^C\bar\zeta^D-16L_{AB}\bar\chi_C\zeta_DL^{CD}\bigg]\\[2mm]
&+\frac{1}{\chi_T^5}\bigg[-6 L_{AB}(\,G\,\bar\zeta_C\zeta_D L^{CD}+\bar G \bar \zeta^C\zeta^D L_{CD})\nonumber\\[2mm]
&-12 L_{aC(A}L_{B)E} L^{CD}\bar\zeta^E\gamma_a\zeta_D+12 L_{C(A}\epsilon_{B)E}L^{DE}\,\tilde h_a\bar\zeta^C\gamma^a\zeta_D\nonumber\\[2mm]
&+6 L_{C(A}\bar\zeta_{B)}\zeta_D \bar \zeta^C\zeta^D+6 \epsilon_{D(A}\epsilon_{B)C}\bar\zeta^C\zeta^D L^{EF}\bar\zeta_E\zeta_F\bigg]\nonumber\\[2mm]
&-\frac{30}{\chi_T^7}L_{AB}\bar\zeta^C\zeta^DL_{CD}\, \bar\zeta_E\zeta_F L^{EF}, \nonumber\\[2mm]
{\cal F}_{T\,ab}&=2D_{[a}\left(\frac{\tilde h_{b]}}{\chi_T}\right)-L_T T_{ab\,AB}\epsilon^{AB}-\bar L_T T_{ab}^{AB}\epsilon_{AB}\nonumber\\[2mm]
&+\frac{1}{\chi_T}\left({{R_{ab}}^C}_BL_{CA}\epsilon^{AB}+\frac12 \left(\bar\zeta^A\rho_{ab}^B\epsilon_{AB}+\bar\zeta_A\rho_{ab\, B}\epsilon^{AB}\right)\right)\nonumber\\[2mm]
&\frac{1}{\chi_T^3}\bigg[-2 L_{[a\,AC}L_{b]\, BD}L^{CD}\epsilon^{AB}+2L_{AB}\bar\zeta_{[a\,C}\gamma_{b]}\zeta^B\epsilon^{AC}\\[2mm]
&-2L_{AB}\bar\zeta_{[a}^B\gamma_{b]}\zeta_C \epsilon^{AC}+L_{[a\, AB}\bar\zeta^B\gamma_{b]}\zeta_C\epsilon^{AC}\bigg]\nonumber\\[2mm]
&+\frac{12}{\chi_T^5}L_{[a\,AB}L^{CD}L_{DE}\epsilon^{DA}\bar\zeta^E\gamma_{b]}\zeta_C. \nonumber
\end{align}
Once the components of the restricted chiral multiplet have been written in terms of the tensor multiplet fields, one obtains the action for the tensor multiplet by using a Lagrangian that expresses the couplings of a vector and a tensor multiplet as follows:
\begin{align}\label{tensorac0}
	S = \int &\left[2 \,B \wedge\left( F_{T}+\frac12 \,L_T\, \bar{\psi}_A \psi_B \epsilon^{AB} + \frac12\,\bar{L}_T\, \bar{\psi}^A \psi^B \epsilon_{AB}\right)\right.  \nonumber \\[2mm]
	&+ \frac{1}{4!} \,e^a e^b e^c e^d \epsilon_{abcd}\, \left(\bar{\lambda}^{T}_A \zeta^A + \bar{\lambda}^{T\, A} \zeta_A- G L_T -  \bar G \bar L_T - L^{AB}Y^{T}_{AB} \right) \nonumber\\[2mm]
	&-\frac{i}{3!} e^a e^b e^c\left( \bar{\psi}^A \gamma_{abc} \zeta_A  L_T+  \bar{\psi}^A \gamma_{abc} \lambda^{T}_A -\bar{\psi}_A \gamma_{abc} \zeta^A  \bar L_T- \bar{\psi}_A \gamma_{abc} \lambda^{T\, A} \right) \nonumber \\[2mm]
	&\left.-\frac{i}{2} e^a e^b \left(\bar{\psi}_A \gamma_{ab} \psi_B L^{AB}  L_T-\bar{\psi}^A \gamma_{ab} \psi^B L_{AB} \bar L_T\right)\right].
\end{align}
One has to use this coupling, rather than directly the action for the vector multiplets because the 2-form $F_T$ already contains second derivatives of the scalar fields and therefore (\ref{tensorac0}) is the right action that gives a two derivative Lagrangian.

As before, we can produce the spacetime Lagrangian by projecting (\ref{tensorac0}) to the  $\theta = d \theta = 0$ surface, whose bosonic part is the following:
\begin{align}\label{tensoraction}
	S_{tensor} = \int d^4 x \,\sqrt{-g} &\left[ \frac{\chi_T}{6}R-\frac{1}{\chi_T} \tilde{H}^\mu \tilde{H}_\mu +\frac{1}{2 \chi_T}D_{\mu}L_{AB} D^{\mu} L^{AB} \right.\nonumber\\[2mm]
	&- \frac{2}{\chi_T} \tilde{H}^\mu \,\omega_\mu{}^C{}_B L_{AC} \epsilon^{AB}+ \frac{1}{\chi_T^3} \frac{\epsilon^{\mu\nu\rho\sigma}}{\sqrt{-g}} B_{\mu\nu} \partial_{\rho} L_{AB} L^{BC} \partial_{\sigma} L_{CD} \epsilon^{CD}\\[2mm]
	&\left.-\frac{G \bar{G}}{\chi_T} +\frac23 \chi_T\,{\cal D}\right].\nonumber
\end{align}

In the previous discussion symmetry arguments fixed the form of the prepotential $\chi_T$ for a single tensor multiplet.
Once on-shell, after dualization of the massless tensor field to a scalar, we expect the resulting scalar $\sigma$-model to reproduce one of the quaternionic-K\"ahler manifolds describing the hypermultiplet self-interactions.
It is clear, though, that a specific form of $\chi_T$ will result in a specific $\sigma$-model.
On the other hand, supersymmetry does not fix uniquely such scalar manifold for a single hypermultiplet.
There are infinite consistent families of manifolds with different numbers of symmetries.
One could therefore ask whether it is possible to generalize this procedure to generate a different ansatz for $\chi_T$.
Since SU(2) invariance and the requirement of the U(1) and Weyl charges for $\chi_T$ uniquely specify it if we use a single tensor multiplet, we could think that the non-linear multiplet could come to our rescue.
As we will see later, the scalar matrix of the non-linear multiplet, which have zero chiral and Weyl charge, are going to be fixed to the identity when used as compensators for the SU(2) superconformal symmetry.
This means that one could use them in $\chi_T$ to generate almost arbitrary functions of the tensor fields that do not respect SU(2) invariance on-shell.
One could, for instance, contract the SU(2) triplet $L^{AB}$ with a constant scalar vector constructed as
\begin{equation}
	m_{AB} = \Phi^i_A \Phi_B^j m_{ij},
\end{equation}
where
\begin{equation}
	m_{ij} = \frac{i}{2} \, (\sigma^r)_i{}^k \epsilon_{kj}\, m^r
\end{equation}
is a triplet of constants, and use this combination to generate a more general expression for $\chi_T$.
Actually, the most general case could have three such independent vectors and the generating potential will be therefore an arbitrary function of these SU(2) invariant contractions, homogeneous of degree 1 in $L_{AB}$.
Once we go to Poincar\`e by fixing $\Phi_i^A = \delta_i^A$, this would lead to a prepotential of the form
\begin{equation}\label{generaltensor}
	\chi_T = |\vec{L}| \ f\left(\frac{L_1}{|\vec{L}|},\frac{L_2}{|\vec{L}|}\right),
\end{equation}
for an arbitrary function $f$.
Unfortunately, when doing so, there is no completion of (\ref{XT0}) by means of fermion bi-linears and/or additional bosonic terms that makes it the lowest component of a vector multiplet.
Thus, this construction produces a single, specific, scalar manifold.

As a final point in this discussion we also note that in (\ref{tensoraction}) there is once more a linear term in the auxiliary ${\cal D}$ field, which would put us on-shell again.
In order to stay off-shell, however, we can add a term proportional to the constraint (\ref{nonlinearconst}) \cite{deWit:1980lyi,Mohaupt:2000mj}, so that we can cancel at the same time the terms proportional to ${\cal D}$ in the tensor action (\ref{tensoraction}) and in the vector action (\ref{vectoraction}).
The bosonic part is given by
\begin{equation}\label{Scon}
	S_{con} = \int d^4 x \sqrt{-g} \, 2 \left({\cal D} - 2 V_b V^b - M_{AB} M^{AB} +\frac14 \Phi_{ai}^B \Phi_B^{ai} - V_a^a\right) (2\chi_V-\chi_T),
\end{equation}
where 
\begin{equation}
	V_a^a = \partial_a V^a +f_a^a+\ldots \,.
\end{equation}
While we are simply adding zero to the Lagrangian, we are effectively replacing the linear term in ${\cal D}$ with quadratic terms in the other auxiliary fields.
A very important fact to note here is that the sum of the vector action (\ref{vectoraction}), tensor action (\ref{tensoraction}) and constraint (\ref{Scon}) not only eliminates the linear terms in ${\cal D}$, but it also reduces the terms proportional to the Ricci scalar and kinetic term for the gravitino to
\begin{equation}
	\int d^4x\,\sqrt{-g} \, \chi_V \left[\frac{R}{2} -  \left(\bar{\psi}_{\mu A}\gamma^{\mu\nu \rho}D_\nu \psi_{\rho}^A + h.c.\right)\right].
\end{equation}

% subsection tensor_multiplet_action (end)

\subsection{The quaternionic manifold} % (fold)
\label{sub:the_quaternionic_manifold}

A massless tensor multiplet in four dimensions is dual to a massless scalar field.
For this reason we can rewrite the action of the tensor multiplet in terms of an action for a hypermultiplet whose scalar $\sigma$-model parameterizes a quaternionic-K\"ahler manifold.
To see this, we need to look at the on-shell action, removing the auxiliary fields $V_a$, $\omega^A{}_B$ and dualizing the tensor field by adding a topological term of the form
\begin{equation}
	S_{top} = -\int H \wedge dw.
\end{equation}

Identifying the dual 1-form of the tensor field strength with $h$ and after integrating by parts the derivative term on the $V^a$ field in (\ref{Scon}), the relevant terms in the Lagrangian can be expressed as a line element, collecting all quadratic terms in $\partial_a L_{AB}$, $V^a$, $h_a$ and $\omega_a^A{}_B$.
Using a vectorial notation for the SU(2) triplets, we can express the geometry of the scalar $\sigma$-model by the line element
\begin{align}\label{linequat}
	ds^2&=- \frac{1}{\chi_T} h^2 - \frac{h}{\chi_T} \left[\frac{L_1}{(L_2)^2+(L_3)^2}(L_2 d L_3 - L_3 dL_2) - \vec{L} \cdot \vec{\omega}\right] - 2 \,V^a  \partial_a{\chi_T}  \nonumber \\[2mm]
	&+ 4(\chi_T - \ell^2)(V^2-|\vec{\omega}|^2) + \frac{1}{4 \chi_T}(d \vec{L} - \vec{\omega} \wedge \vec{L})^2 + h dw,
\end{align}
where $\ell^2 = 2 \chi_V$, which is going to be related to the size of the scalar manifold, as will be clear in the following.
Integrating out the auxiliary fields $\vec{\omega}$ and $V^a$ and the field strangth $h$, we are left with the metric of the corresponding quaternionic-K\"ahler manifold, depending only on $\vec{L}$ and $w$.
It is easy to recognize this manifold if we introduce spherical coordinates for $\vec{L}$, namely $L_1 =\ell^2\, \rho \cos \theta$, $L_2 =\ell^2\, \rho \sin \theta\, \sin \phi$, $L_3 =\ell^2\, \rho\, \sin \theta\, \cos \phi$:
\begin{equation}\label{lineq2}
	ds^2 =-\frac{\ell^2}{4}\left[ \frac{d \rho^2}{\rho(\rho -1)} + \rho(\rho-1)\left(dw^2 + d \theta^2 + d \phi^2 + 2 \cos\theta\, dw d \phi\right)\right].
\end{equation}
This is indeed the metric of the quaternionic manifold SO(4,1)/SO(4), with the correct signature for the kinetic terms whenever $\rho > 1$.
It is suggestive to rewrite this metric as
\begin{equation}
	ds^2 = -\frac{\ell^2}{4}\left[\frac{dr^2}{r(1-r)^2} + \frac{4r}{(1-r)^2} d \Omega_{S^3}^2\right],
\end{equation}
by setting $\rho = 1/(1-r)$, which is the metric for Euclidean AdS space, seen as the hypersurface
\begin{equation}
	- (X_0)^2 + (X_1)^2 + (X_2)^2 + (X_3)^2 + (X_4)^2 = -\frac{\ell^2}{4},
\end{equation}
embedded in Minkowski space in five dimensions, using the solution
\begin{align}
	X^0 &= \frac{\ell}{2} \left(\frac{1+r}{1-r}\right), \\[2mm]
	X^I &= \ell \,\frac{\sqrt{r}}{1-r} \Omega^I,
\end{align}
where $\Omega^I$ is the angular parametrization of the 3-sphere.
The $S^3$ metric has been written here using its Hoof-fibration structure, which is going to be useful in the following.
This is made explicit by introducing the triplet of SU(2) left-invariant forms
\begin{align}
		\hat{\sigma}_1 + i\, \hat{\sigma}_2 =& -\frac12 e^{i w} \left(d \theta -i\,\sin \theta d \phi\right),\\[2mm]
	\hat{\sigma}_3 =& \frac12 (dw +\cos \theta d \phi),
\end{align}
which satisfy $d \hat{\sigma}_i = \epsilon_{ijk} \hat{\sigma}_j \wedge \hat{\sigma}_k$, so that $d \Omega_{S^3}^2 = \sum_i (\hat{\sigma}_i)^2$.

One should have expected this result because of the very high symmetry of the prepotential function $\chi_T$.
Less symmetric $\sigma$-models could follow from less symmetric $\chi_T$ functions, if one could complete consistently (\ref{XT0}) to the lowest component of a reduced chiral multiplet.

% subsection the_quaternionic_manifold (end)

\subsection{Other multiplet relations} % (fold)
\label{sub:other_multiplet_relations}

In the previous part of the work we encountered examples of maps between different superconformal multiplets. 
In section~\ref{sub:chiral_multiplet}, in fact, we showed how to build a chiral multiplet out of a product of vector multiplets, while in section \ref{sub:tensor_multiplet} a tensor multiplet has been constructed as a function of the components of the vector multiplet. 
While it revealed to be extremely useful in the construction of superconformal actions, this fact can be of interest in its own right and we would like to elaborate it further in what follows.

Let us notice first of all that, in the same spirit of section \ref{sub:tensor_multiplet}, one can try to construct a vector multiplet out of a non-linear multiplet. 
The starting point is always the identification of the component $L$ of the vector. 
In this case the ansatz
\begin{equation}
L_{NL} = M^{AB}\epsilon_{AB}
\end{equation} 
has the correct Weyl and chiral weights. 
From the transformation rules of $M^{AB}$, one can check that $DL_{NL}$ organizes exactly in the form (\ref{DX}), with
\begin{align}
\lambda^{NL}_A &= 4 \epsilon_{AB}\chi^B - \gamma^{ab}\tau_A T_{ab}^{BC}\epsilon_{BC} - 4 \epsilon_{AB}\gamma^a \tau^B V_a \\[2mm]
&- 2 \tau_A M^{BC}\epsilon_{BC}-2 \epsilon_{AB}\gamma^a\tau_a^B - 2\epsilon_{AB}\Phi_i^B \gamma^a \Phi_{aC}^i\tau^C.
\end{align}
Continuing with this procedure one can check that also the other components correctly fulfill the Bianchi identities.
Unfortunately, higher compontents contain terms with second derivatives of the fields, as can be seen from the inspection of their bosonic part
\begin{align}
{\cal F}_{ab}^{NL} &= -i (\mathcal{F}^+_{ab}-\mathcal{F}^-_{ab}) - 2 M_{CD}T^{CD}_{ab}-2M^{CD}T_{ab \, CD}-4 V_{[ab]},\\[2mm]
Y^{NL}_{AB} &= - 4 \epsilon_{(A|C|}\Phi_i^C \Phi_{aB)}^i V_a - \epsilon_{(A|C|}\Phi_i^C D^a \Phi_{aB)}^i.
\end{align}
This implies that, once inserted in the superconformal action for the vector multiplet (\ref{vectoraction}), one gets higher derivative couplings, which are an undesirable feature if we focus on the Poincar\'e low energy action, but could be of interest in the construction of more general conformal models.

Another interesting fact to be explored would be the construction of a chiral multiplet as a product of vector multiplets, in which one or more of the latter are expressed as function of tensor multiplets. Again this is going to produce an invariant action with the presence of higher derivatives and, for this reason, we did not investigate it further.

Finally we think it would be important to investigate also the existence of the inverse for any of the aforementioned maps, but this is beyond the actual purpose of the present work.

% subsection other_multiplet_relations (end)

\subsection{The Poincar\'e limit} % (fold)
\label{sub:the_poincar_e_limit}

The analysis of the conformal couplings of vector and tensor multiplets we presented so far has not been done to study these models in a (super)conformal invariant framework, but rather to provide a derivation of the off-shell formulation of Poincar\'e supergravity coupled to matter.
This means that we have to think of the conformal (super)symmetries as a mere tool to construct easily the Lagrangian and supersymmetry rules of the Poincar\'e theory.
In fact, by means of the additional symmetries present in the superconformal group we can linearly realize some of the symmetries and simplify some of the couplings that otherwise would be complicated to determine.
The price we have to pay is to introduce a certain number of fields that act as compensators for such symmetries and whose fixing to a definite value brings us back to the ordinary Poincar\'e theory.
We need to fix fields on which special conformal transformations $K_a$, dilatations $D$, conformal supersymmetries $S^A$, chiral U(1) $T_{U(1)}$ and SU(2) $T^A{}_B$ transformations act as gauge symmetries.
While doing so, we have to be sure that the conditions we impose remain invariant with respect to the residual super-Poincar\'e group.
This happens if we redefine the Poincar\'e supersymmetries by additional field-dependent transformations of the symmetries we fixed.
Clearly in the following discussion all expressions have to be understood in ordinary spacetime and not in superspace.

Following \cite{Mohaupt:2000mj}, we start by fixing invariance under dilatations, choosing a condition that brings the term proportional to the Ricci tensor to the usual Einstein--Hilbert form
\begin{equation}
\label{Dfix}
	\chi_V=-i\left(L^\Lambda \bar P_\Lambda - \bar L^\Lambda P_\Lambda\right) = \kappa^{-2}.
\end{equation}
We restored in this formula the dimensionful coupling constant $\kappa$, which explicitly breaks conformal invariance, but in the following we will use natural units.
This constraint is not invariant under dilatations nor supersymmetry, but we can solve this by requiring that
\begin{equation}
	\delta_{Q}^{Poincar\acute{e}}(\epsilon) \chi_V = \delta_D(\epsilon) \chi_V + \delta_Q(\epsilon) \chi_V = 0,
\end{equation}
i.e.~that Poincar\'e supersymmetry transformations include field-dependent dilatation transformations such that the above expression vanishes.
However, having non-zero $\Lambda_D$ would affect the transformation rule of the vielbein, because
\begin{equation}
	\delta_D e^a = - \Lambda_D e^a
\end{equation}
and this would mean that we would need to perform non-trivial field redefinitions on the metric to go back to the standard form of the graviton supersymmetry transformation
\begin{equation}
	\delta_Q^{Poincar\acute{e}} e^a = \bar{\epsilon}^A \gamma^a \psi_A - \bar{\psi}^A \gamma^a \epsilon_A.
\end{equation}
For this reason we fix $\delta_Q \chi_V = 0$, which, in turn, implies $\Lambda_D = 0$.
This can be obtained by setting 
\begin{equation}
\label{Sfix}
	\left({\bar P}_{\Lambda} - {\bar L}^\Sigma P_{\Lambda \Sigma}\right) \lambda^\Lambda_A = -2i\, (\Im P)_{\Lambda \Sigma}\bar{L}^\Sigma \lambda^\Lambda_A = 0,
\end{equation}
which now fixes one fermion as a function of the others and breaks conformal supersymmetry.
In fact this condition does not remain invariant under supersymmetry nor conformal supersymmetry.
We restore Poincar\'e supersymmetry invariance by requiring that the conformal supersymmetry parameter $\eta^A$ be a function of the fields as follows
\begin{align}
	\eta_A =& -\frac{i}{2}\bigg[\bar\epsilon_B \lambda^{B\,\Lambda}(\bar P_{\Lambda\Sigma}-P_{\Lambda\Sigma})\lambda^\Sigma_A - P_{\Lambda\Sigma\Gamma}\bar\epsilon^B\lambda_B^\Gamma\,\bar L^\Lambda \lambda_A^\Sigma \nonumber\\[2mm]
&+ \left(\bar P_\Lambda - P_{\Lambda\Sigma}\bar L^\Sigma \right)\left(\gamma^a \epsilon_A L_a^\Lambda + \frac12 \slashed{\cal F}^\Lambda \epsilon_{AB}\epsilon^B+Y_{AB}^\Lambda\epsilon^B\right) \bigg].
\end{align}
We can then move safely to special conformal transformations, which are fixed by the condition
\begin{equation}
\label{Kfix}
	b = 0,
\end{equation}
and the requirement that 
\begin{equation}
	\Lambda^a_K =\frac12\left(\bar \epsilon^A \phi^a_{A}+ \bar\epsilon_A\phi^{aA} -\bar\eta^A \psi^a_{A}-\bar\eta_A\psi^{aA} -\bar\epsilon^A\gamma^a\chi_A-\bar\epsilon_A\gamma^a\chi^A \right).
\end{equation}
Finally we can fix SU(2) invariance by choosing
\begin{equation}
\label{SU2fix}
	\Phi^A_i = \delta^A_i
\end{equation}
and at the same time requiring that the SU(2) parameter be fixed in terms of supersymmetry as
\begin{equation}
	\Lambda^A_B\,\delta^B_i =2 \bar\epsilon_i \tau^A - \delta_i^A \bar\epsilon_C \tau^C - 2\bar\epsilon^A \tau_i + \delta_i^A \bar\epsilon^C \tau_C.
\end{equation}
Note that although we fixed local SU(2) invariance, our actions are still invariant under global SU(2) transformations.
We still have local U(1) invariance, which is usually fixed by imposing
\begin{equation}
	X^0 = \bar{X}^0,
\end{equation}
removing another scalar field from the vector multiplet content.
Therefore, the resulting theory has only $n_V$ vector multiplet complex scalars $z^i$ ($i=1,\ldots,n_V$) and $n_V$ vector multiplet doublets of fermions, as a consequence of (\ref{Sfix}).

We can now see explicitly how this procedure affects the action, written in terms of the component fields.
We focus on the vector multiplet action and the constraint, because they are the ones that are most affected by this procedure.

After a very long and tedious calculation and up to a total derivative term, the vector multiplet action  can be written as the sum of four pieces, collecting together the kinetic terms, the terms depending on the auxiliary fields, the Pauli-like couplings and the 4-fermi interactions, which we discard in the following:
\begin{equation}
	\label{gaugefixed}
		L_{vector} = L_{kin} + L_{Pauli} + L_{aux}.
\end{equation}
Up to 4-fermi interactions we therefore have
\begin{equation}
	\begin{split}
		e^{-1}L_{kin} &= \chi_V\left(\frac{R}{6} - \frac43\, {\cal D}\right)  - 2 (\Im P)_{\Lambda \Sigma} \, D^a L^\Lambda D_a \bar{L}^\Sigma+\left[ i\, P_{\Lambda \Sigma}\, {F}^{+\Lambda}_{ab} {F}^{ab + \Sigma} \right. \\[2mm]
		& -\frac13\, \chi_V \,\bar{\psi}_{a A} \gamma^{abc} D_b \psi_c^A  + (\Im P)_{\Lambda \Sigma}{L}^\Sigma\, \bar{\psi}_{aA} \gamma^{abc} \psi_c^A D_b \bar L^\Lambda \\[2mm]
		&- (\Im P)_{\Lambda \Sigma}\, \bar \lambda_A^\Lambda \gamma^a D_a\lambda^{A\,\Sigma} -\frac{i}{2}\, P_{\Lambda \Sigma \Gamma} \,D_a L^\Lambda\, \bar{\lambda}_A^\Sigma \gamma^a \lambda^{A \Gamma} \\[2mm]
	&\left.-\frac43 (\Im P)_{\Lambda \Sigma} \,\bar{L}^{\Sigma} \bar{\lambda}^{\Lambda}_A \gamma^{ab} D_a \psi_b^A +2\,(\Im P)_{\Lambda \Sigma}\,\bar{\psi}_a^A \slashed{D}\bar{L}^\Lambda \gamma^a \lambda_A^\Sigma + h.c.\right],\\[2mm]
\end{split}
\end{equation}
\begin{equation}
	\begin{split}
		e^{-1}L_{Pauli} &=4 (\Im P)_{\Lambda \Sigma} {F}^{-ab \Lambda} \bar{\psi}_{a A} \gamma_b \lambda_B^\Sigma \epsilon^{AB} + 4 (\Im P)_{\Lambda \Sigma} \bar{\psi}_{aA} \psi_{b B} \epsilon^{AB}\, {F}^{-ab \Sigma} L^\Lambda \\[2mm]
		&-\frac{i}{4}\,P_{\Lambda \Sigma \Gamma} \bar{\lambda}^\Lambda_C \gamma^{ab} \lambda_D^\Sigma \epsilon^{CD} \,{F}_{ab}^\Gamma + h.c.\ ,
	\end{split}
\end{equation}
and
\begin{equation}
	\begin{split}
		e^{-1}L_{aux} &= (\Im P)_{\Lambda \Sigma} \, Y_{AB}^\Lambda Y^{AB\, \Sigma} + \left[ 4\, (\Im P)_{\Lambda \Sigma}\, \bar{L}^\Lambda \,{F}^{+ab \Sigma} T_{ab}^{CD} \epsilon_{CD}\right. \\[2mm]
		&- 2\,(\Im P)_{\Lambda \Sigma} L^\Lambda L^\Sigma\, T_{ab\,AB} \epsilon^{AB} T^{ab}_{CD} \epsilon^{CD}   -\frac{16}{3}\, (\Im P)_{\Lambda \Sigma}L^\Lambda \,\bar{\chi}_A \lambda^{A \Sigma} -\frac53\, \chi_V \,\bar{\psi}_{a A} \gamma^a \chi^A \\[2mm]
	&-\frac43 (\Im P)_{\Lambda \Sigma} \, \bar{\lambda}^{\Lambda A} \gamma^a \psi^{bB}\, \epsilon_{AB}\, T_{abCD} {L}^\Sigma \epsilon^{CD}  -4\,(\Im P)_{\Lambda \Sigma} L^\Lambda L^\Sigma\, \bar{\psi}_{a A} \psi_{bB} \epsilon^{AB}\, T^{ab}_{CD} \epsilon^{CD} \\[2mm]
	&-4\, (\Im P)_{\Lambda \Sigma} \, \bar{\lambda}^\Lambda_A \gamma^a \psi^b_B \epsilon^{AB} \, T_{abCD} {L}^\Sigma \epsilon^{CD}  +\frac53\, \chi_V \, \bar{\psi}_{a A} \psi_{bB} \epsilon^{AB} \, T^{ab CD} \epsilon_{CD}\\[2mm]
		& +\frac{i}{2}\, P_{\Lambda \Sigma \Gamma} \, Y^{AB \Lambda} \bar{\lambda}^\Sigma_A \lambda_B^\Gamma - i P_{\Lambda \Sigma}\, Y^{\Lambda AB} \, \bar{\psi}_{aA}\gamma^a \lambda^{\Sigma}_B + \\[2mm]
		&\left.- i P_{\Lambda}\, \bar{\psi}_{aA}\gamma^{ab}\psi_{bB} Y^{\Lambda AB}+ h.c. \right]. \\[2mm]
	\end{split}
\end{equation}
The constraint action, up to 4-fermion couplings, can be written in terms of the component fields as follows
\begin{equation}
	\begin{split}
		e^{-1}L_{con} &= 4 \chi_V \left[\frac13 {\cal D} + \frac{1}{12} R + \frac14 D_{a}\Phi^A{}_i D^a \Phi_A^i -2 V_aV^a - M_{AB} M^{AB} -D^a V_a  \right.\\[2mm]
		&+ \left(\bar{\tau}_A \gamma^a D_a \tau^A- \frac{1}{6} \, \bar{\psi}_{aA}\gamma^{abc}D_b \psi_c^A +\frac16 \bar{\tau}^A \gamma^{ab} D_a \psi_{bA} - \frac12\bar{\psi}_{aA} \gamma^{ab} D_b \tau^A\right. \\[2mm]
		&+\frac16 \bar{\psi}_{aA} \gamma^a \chi^A-\frac43 \bar{\tau}^A \chi_A+ \bar{\psi}_a^A \gamma^a \psi^b_A \,V_b  -2 \bar{\tau}^A \psi_{aA} V^a+ \frac12\,\bar{\psi_a}^A \gamma^{ab} \psi_{b}^B M_{AB}  \\[2mm]
		& + \frac14 \, \Phi_B^i D_a \Phi^A_i \, \bar{\psi}_b^B \gamma^{bc}\gamma^a \psi_{cA}-\bar{\tau}^A \gamma^{ab}\psi_{aB}  \Phi_A^i D_b \Phi^B_i \\[2mm]
		&- \bar{\psi}_{aA} \tau^B\, \Phi_B^i\, D^a \Phi^A_i  + \bar{\tau}^A \gamma^a \tau_B\, \Phi_A^i \, D_a \Phi_i^B  \\[2mm]
		&\left.\left. -\bar{\tau}_A \slashed{T}^{AB} \tau_B -\frac13 \bar{\tau}_A \slashed{T}^{AB} \gamma^a \psi_{aB} + \frac23 \bar{\psi}_{aA} \psi_{bB} T^{ab AB} + h.c.\right)\right], \\[2mm]
	\end{split}
\end{equation}
where we set $\chi_T = 0$ for the time being.

We can now proceed by implementing the constraints that produce the Poincar\'e supergravity action.
We can solve (\ref{Dfix}) by introducing the so-called symplectic projective covariant sections $L^\Lambda$ and $M_{\Lambda}$ of \cite{Andrianopoli:1996cm}.
In detail, from (\ref{Dfix}) one deduces that 
\begin{equation}
	L^\Lambda = e^{K/2} X^\Lambda, \quad P_{\Lambda} = M_{\Lambda} = e^{K/2} F_{\Lambda},
\end{equation}
where
\begin{equation}
	K = - \log [-i(X^\Lambda \bar{F}_{\Lambda} - \bar{X}^\Lambda F_{\Lambda})]
\end{equation}
is the K\"ahler potential of the vector multiplet scalar manifold and $F(X) = e^{-K}P(L)$ is the special-K\"ahler prepotential, holomorphic in the $X^\Lambda$ fields.
This also implies $P_{\Lambda \Sigma} = F_{\Lambda \Sigma}$, $P_{\Lambda \Sigma \Gamma} = e^{-K/2} F_{\Lambda \Sigma \Gamma}$ and $P_{\Lambda \Sigma \Gamma \Delta} = e^{-K} F_{\Lambda \Sigma \Gamma \Delta}$.
Although the full special-K\"ahler geometry can be recovered only on-shell, as it is going to be clear momentarily, we introduce all the special-K\"ahler quantities from the beginning so that it is easier to recover the on-shell result.

Once we removed the extra complex scalar field, we can use the condition (\ref{Sfix}) to remove the corresponding fermion fields.
It is indeed easy to see that the definition
\begin{equation}
	\lambda_A^\Lambda = f_i^\Lambda \, \lambda^i_A
\end{equation}
solves identically the constraint, where the derivatives $f_i^\Lambda \equiv D_i L^\Lambda$ act as a projector to the $n_V$-dimensional subspace of gaugini.
This expression allows further simplification of the fermion expressions in the Lagrangian, also using the identities 
\begin{align}
	e^{K/2} C_{ijk} &= F_{\Lambda \Sigma \Gamma} f_i^\Lambda f_j^\Sigma f_k^\Gamma, \\[2mm]
	(\Im F)_{\Lambda \Sigma} D_i f_j^\Lambda &= \frac{i}{2} \, e^{-K/2}\, F_{\Lambda \Sigma \Gamma} \, f_i^\Sigma f_j^\Gamma, \\[2mm]
	f_i^\Lambda f_j^\Sigma f_k^\Gamma f_l^\Delta F_{\Lambda \Sigma \Gamma \Delta} &= e^{K} \left[D_l C_{ijk} - 6 i\, C_{lim} C_{jkn} g^{m \bar{m}} g^{n \bar n} \bar{f}_{\bar m}^\Lambda \bar{f}_{\bar n}^\Sigma (\Im F)_{\Lambda \Sigma}\right],
\end{align}
where we used the special geometry relations
\begin{align}
	&V \equiv (L^\Lambda, M_{\Lambda}), \qquad U_i \equiv D_i V = \left(\partial_i + \frac12 \partial_i K\right) V = (f_i^\Lambda , h_{i \Lambda}), \\[2mm]
	&\langle V, \bar{V} \rangle \equiv \bar{L}^\Lambda M_{\Lambda} - L^\Lambda \bar{M}_{\Lambda} = -i, \\[2mm]
	&\langle V, U_i \rangle  = \langle {\bar V}, U_i \rangle = 0, \\[2mm]
	& \langle U_i ,  \bar{U}_{\bar \jmath} \rangle = i\, g_{i \bar \jmath}, \\[2mm]
	& D_i U_j = i\, C_{ijk} g^{k \bar k} \bar{U}_{\bar k}, \\[2mm]
	& U^{\Lambda \Sigma} \equiv f_i^\Lambda g^{i \bar \jmath} \bar{f}_{\bar \jmath}^\Sigma = \frac12 \, (\Im F)^{\Lambda \Sigma} + L^\Lambda \bar{L}^\Sigma. \label{Udef}
\end{align}

At this point, combining the actions together, we see that the linear terms in ${\cal D}$ disappear, while the kinetic terms of graviton and gravitino and the coupling of the auxiliary field $\chi^A$ with the gravitino simplify to the expected form (again up to total derivatives and discarding 4-fermion couplings) 
\begin{equation}
	L_{final} = L_{kin} + L_{Pauli} + L_{aux},
\end{equation}
where
\begin{equation}
		\begin{split}
			e^{-1}L_{kin} &= \frac{R}{2} - g_{i \bar\jmath}\nabla^a z^i \nabla_a \bar{z}^{\bar\jmath}+ \left({\cal A}_a -\frac{i}{2} \nabla_a z^i \partial_i K + \frac{i}{2} \nabla_a \bar{z}^{\bar \jmath} \partial_{\bar\jmath} K \right)^2 \\[2mm]
			& +\left[ i\, F_{\Lambda \Sigma} \,{F}^{+\Lambda}_{ab} {F}^{ab + \Sigma}-\bar{\psi}_{a A} \gamma^{abc} \left(\nabla_b +\frac14\,\nabla_a z^i \partial_i K - \frac14 \nabla_a \bar{z}^{\bar \jmath} \partial_{\bar\jmath} K\right) \psi_c^A \right.   \\[2mm]
			&- \frac12\,g_{i \bar\jmath}\, \bar \lambda_A^i \gamma^a \left(\nabla_a - \frac{i}{2}{\cal A}_a -\frac12 \nabla_a z^k \partial_k K + \frac12 \nabla_a \bar{z}^{\bar k} \partial_{\bar k} K \right)\lambda^{A\,\bar\jmath}  \\[2mm]
			&+ 4\, \bar{\tau}_A \gamma^a \left(\nabla_a-\frac{i}{2} {\cal A}_a\right) \tau^A+\frac83\, \bar{\tau}_A \gamma^{ab} \left(\nabla_a-\frac{i}{2} {\cal A}_a\right) \psi_{b}^A \\[2mm]
			&\left.+ g_{i \bar\jmath} \, \bar{\psi}_a^A \slashed{\nabla}\bar{z}^{\bar\jmath} \gamma^a \lambda_A^i+ h.c.\right],\\[2mm]
	\end{split}
\end{equation}
\begin{equation}
	\begin{split}
		e^{-1}L_{Pauli} &=  -\frac12\,(\Im F)_{\Lambda \Sigma} \, D_i f_j^\Sigma \,\bar{\lambda}^i_C  \gamma^{ab} \lambda_D^j\, \epsilon^{CD} \,{F}_{ab}^\Gamma \\[2mm]
	&+4 (\Im F)_{\Lambda \Sigma} f_i^\Sigma\, {F}^{-ab \Lambda} \,\bar{\psi}_{a A} \gamma_b \lambda_B^i\, \epsilon^{AB} + 4 (\Im F)_{\Lambda \Sigma}L^\Lambda \, \bar{\psi}_{aA} \psi_{b B} \epsilon^{AB}\, {F}^{-ab \Sigma} + h.c.\ ,
	\end{split}
\end{equation}
and
\begin{equation}
	\begin{split}
		e^{-1}L_{aux} &= (\Im F)_{\Lambda \Sigma} \, Y_{AB}^\Lambda Y^{AB\, \Sigma}  -8 V_aV^a - 4M_{AB} M^{AB}- \omega_a{}^A{}_B \, \omega^a{}_A{}^B\\[2mm]
	& \left[ 4\, (\Im F)_{\Lambda \Sigma}\, \bar{L}^\Lambda \,{F}^{+ab \Sigma} T_{ab}^{CD} \epsilon_{CD}- 2\,(\Im F)_{\Lambda \Sigma} L^\Lambda L^\Sigma\, T_{ab\,AB} \epsilon^{AB} T^{ab}_{CD} \epsilon^{CD}    \right. \\[2mm]
	&+3\,  \bar{\psi}_{a A} \psi_{bB} \epsilon^{AB} T^{ab CD} \epsilon_{CD} -4\, (\Im F)_{\Lambda \Sigma} f_i^\Lambda {L}^\Sigma\, \bar{\lambda}^i_A \gamma^a \psi^b_B \epsilon^{AB} T_{abCD} \epsilon^{CD}  \\[2mm]
	& -4\,(\Im F)_{\Lambda \Sigma} L^\Lambda L^\Sigma\, \bar{\psi}_{a A} \psi_{bB} \epsilon^{AB} T^{ab}_{CD} \epsilon^{CD} \\[2mm]
	&- i F_{\Lambda \Sigma}\, f^\Sigma_i \, Y^{\Lambda AB} \bar{\psi}_{aA} \gamma^a \lambda_B^i - i M_{\Lambda} \bar{\psi}_{aA} \gamma^{ab} \psi_{bB} Y^{\Lambda AB} \\[2mm]
	&-\frac43\,\bar{\tau}^A \gamma^{ab}\psi_{aB} \, \omega_b{}^B{}_A -4 \bar{\psi}_{aA} \tau^B\, \omega^{aA}{}_B   -\frac12\, g_{i\bar\jmath} \,\bar{\lambda}^i_A \gamma^a \lambda^{B \bar\jmath} \omega_a{}^A{}_B\\[2mm]
		&-\bar{\psi}_{a A} \gamma^a \chi^A  +(\Im F)_{\Lambda \Sigma} D_i f_j^\Sigma\,  Y^{AB \Lambda} \left(\bar{\lambda}^i_A \lambda_B^j\right)  \\[2mm]
							&-\frac{16}{3} \bar{\tau}^A \chi_A +4\, \bar{\psi}_a^A \gamma^a \psi^b_A \,V_b  -8 \bar{\tau}^A \psi_{aA} V^a+ 2\bar{\psi_a}^A \gamma^{ab} \psi_{b}^B M_{AB}\\[2mm]
							&\left.\left. -4\bar{\tau}_A \slashed{T}^{AB} \tau_B -\frac43 \bar{\tau}_A \slashed{T}^{AB} \gamma^a \psi_{aB} + h.c.\right)\right], \\[2mm]
	\end{split}
\end{equation}
Here $\nabla$ contains the spin connection and the Levi--Civita connection of the scalar $\sigma$-model.
Note that the U(1) connection ${\cal A}_a$ on-shell contains the K\"ahler connection, providing the appropriate structures required by Special K\"ahler geometry.

% subsection the_poincar_e_limit (end)

% section actions (end)

\section{Gauging} % (fold)
\label{sec:gauging}

In this last section we are going to discuss a gauging of the previous action.
There are various possibilities that can be considered and are compatible with conformal invariance. 
In our case we would like to see if it is possible to generate a coupling of the tensor field to the vectors as the on-shell result of a modification of the actions discussed above, by means of new off-shell couplings.
In particular, we are interested in gaugings involving the tensor field, because it is known that one cannot obtain all possible gaugings using the superconformal construction of Poincar\`e supergravity without them.
We showed in previous sections that the vector multiplet couplings are fixed in terms of a prepotential function $P(L)$, which is related to the Special K\"ahler geometry prepotential $F(X)$ by a rescaling.
Unfortunately the latter exists only in special symplectic frames. 
This implies that the gaugings one can obtain using only vector fields starting from such prescription are limited.
The way out is to use tensor fields, which are usually introduced as auxiliary fields related by additional duality relations to the other fields \cite{deVroome:2007unr,deWit:2011gk}.
In the following we want to do it in a new way that involves the full tensor multiplet and preserves the off-shell structure of the theory.
We also want to make use of the non-linear multiplet as compensator rather than using a hypermultiplet like in \cite{deVroome:2007unr,deWit:2011gk}.

A natural proposal is to add the superconformal invariant action specifying the couplings of a tensor multiplet to vector multiplets:
\begin{align}\label{acttens}
	S = e_{\Lambda}\int &\left[2 \,B \wedge\left( F^{\Lambda} +\frac12\, L^\Lambda \, \bar{\psi}_A \psi_B \epsilon^{AB} + \frac12\, \bar{L}^\Lambda\, \bar{\psi}^A \psi^B \epsilon_{AB}  \right)\right.\nonumber\\[2mm]
	&\left.- \frac{1}{4!} \,e^a e^b e^c e^d \epsilon_{abcd}\, \left( G L^\Lambda+ \bar G \bar L^\Lambda + L^{AB}Y^{\Lambda}_{AB} - \bar{\lambda}^{\Lambda}_A \zeta^A -  \bar{\lambda}^{\Lambda A} \zeta_A\right)\right. \\[2mm]
	&-\frac{i}{3!} e^a e^b e^c\left( \bar{\psi}^A \gamma_{abc} \zeta_A  L^{\Lambda}+  \bar{\psi}^A \gamma_{abc} \lambda^{\Lambda}_A -\bar{\psi}_A \gamma_{abc} \zeta^A  \bar L^{\Lambda}- \bar{\psi}_A \gamma_{abc} \lambda^{\Lambda A} \right) \nonumber \\[2mm]
	&\left.-\frac{i}{2} e^a e^b \left(\bar{\psi}_A \gamma_{ab} \psi_B L^{AB}  L^\Lambda-\bar{\psi}^A \gamma_{ab} \psi^B L_{AB} \bar L^\Lambda\right)\right].\nonumber
\end{align}
We already used this action to construct the self-interactions of a tensor multiplet and its couplings to gravity. 
In this case the action should be taken literally and it expresses the couplings of a tensor multiplet to a certain number of vector multiplets specified by the charges $e_{\Lambda}$.
To see that we obtain the desired effect, we focus on the bosonic sector of the spacetime section of this action
\begin{equation}
	S_m = \int d^4 x \,\sqrt{-g}\, \left[-L^{AB} Y_{AB}^\Lambda e_{\Lambda} - \left( G L^\Lambda e_{\Lambda} - i B_{ab}^+ F^{ab \Lambda +} e_{\Lambda} + h.c.\right)  \right]
\end{equation}
and we consider its addition to the system described by (\ref{vectoraction}), (\ref{tensoraction}) and (\ref{Scon}).

The first thing to note is the coupling between the vector and tensor fields:
\begin{equation}
	\frac12 \, \epsilon^{abcd} B_{ab} F_{cd}^\Lambda e_{\Lambda}.
\end{equation}
An integration by parts of this coupling generates a new term in the dual of the tensor field-strength coupled to a linear combination of the vector fields specified by the charges $e_{\Lambda}$.
This is therefore a new term in (\ref{linequat}) of the form
\begin{equation}
	+2\tilde{H}^a A^\Lambda_a e_{\Lambda},
\end{equation}
which changes the equation of motion for the $\tilde{H}$ field and, once also the other auxiliary fields have been integrated out, results in a change of the line element (\ref{lineq2}) to
\begin{equation}
	ds^2 =- Dq^u h_{uv} Dq^v = -\frac{\ell^2}{4}\left[ \frac{d \rho^2}{\rho(\rho -1)} + 4\rho(\rho-1)\left(\hat{\sigma}_1^2+\hat{\sigma}_2^2 + (\hat{\sigma}_3 + e_{\Lambda}A^\Lambda)^2\right)\right],
\end{equation}
where $q^u =(\rho, w, \theta, \phi)$ denote the hypermultiplet scalar fields.
This signals the electric gauging of the shift symmetry of the angular variable $w$, dual to the tensor field.
In fact, we can put the whole lagrangian in the form of a standard electric gauging of $N=2$ supergravity, writing 
the Poincar\'e invariant bosonic lagrangian as \cite{Andrianopoli:1996cm}
\begin{align}\label{finalac}
	S_{tot} = \int d^4 x\, \sqrt{-g}& \left[\frac{R}{2} - g_{i \bar{\jmath}} \partial_\mu z^i \partial^\mu \bar{z}^{\bar{\jmath}} - h_{uv} D^\mu q^u D_\mu q^v -V(q,z,\bar{z}) \right. \nonumber\\[2mm]
	&+\left. {\cal I}_{\Lambda \Sigma} F^{\Lambda}_{\mu\nu} F^{\Sigma \mu\nu} - \frac12 {\cal R}_{\Lambda \Sigma} \frac{\epsilon^{\mu\nu \rho \sigma}}{\sqrt{-g}} F_{\mu\nu}^\Lambda F_{\rho \sigma}^\Sigma\right],
\end{align}
where one can see the appearance of a non-trivial scalar potential.
The gauge kinetic couplings of the vector fields in (\ref{finalac}) have been obtained by integration of the auxiliary fields $T_{ab}^{AB}$, so that one recovers the match with the standard formulation of $N=2$ supergravity
\begin{equation}
	{\cal N}_{\Lambda \Sigma} = R_{\Lambda \Sigma} + i \,{\cal I}_{\Lambda \Sigma} = \bar{F}_{\Lambda \Sigma} - 2i\, \bar{T}_{\Lambda} \bar{T}_{\Sigma} (L \,\Im F\, L),
\end{equation}
where
\begin{equation}
	T_{\Lambda} = -i \frac{(\Im F)_{\Lambda \Sigma}\bar{L}^\Sigma}{\bar{L}^\Delta (\Im F)_{\Delta \Gamma}\bar{L}^\Gamma}.
\end{equation}
Also, the integration of the auxiliary fields $Y^\Lambda_{AB}$ and $G$ gives the scalar potential
\begin{equation}
	V = \frac18\, \chi_T^2\, e_{\Lambda} (\Im F)^{\Lambda \Sigma} e_{\Sigma} - \chi_T |L^\Lambda e_{\Lambda}|^2.
\end{equation}
This can be brought to the standard supergravity form by properly identifying the quaternionic-K\"ahler structure of the scalar manifold. 
For $\ell^2 = 2 \chi_V = 2$, our scalar model is equivalent to the one in \cite{Behrndt:2001km} with $n=-1$, $\kappa =4$, $\zeta = e^{-i\,\phi}\tan(\theta/2)$ and $\tau = - (w+\phi)/2$, so that one recognizes the quaternionic connection 
\begin{align}
	\omega^1 =& - \frac{\sqrt{\rho-1}}{\sqrt{2 \rho}} \, i \, (u-\bar{u}), \\[2mm]
	\omega^2 =& - \frac{\sqrt{\rho-1}}{\sqrt{2 \rho}} \, (u+\bar{u}), \\[2mm]
	\omega^3 =& - \frac{\rho}{\sqrt{2(\rho-1)}} \,i\,(v-\bar{v})+ (1+\cos \theta) d \phi, 
\end{align}
where
\begin{align}
	u =& - \sqrt2 \sqrt{\rho(\rho-1)} \,e^{i (\phi-w)}(\hat{\sigma}_1 - i\, \hat{\sigma}_2), \\[2mm]
	v =& \frac{d \rho}{\sqrt2 \sqrt{\rho(\rho-1)}} - i\, \sqrt2 \sqrt{\rho(\rho-1)}\; \hat{\sigma}_3.
\end{align}
This produces the SU(2) curvatures 
\begin{align}
	\Omega^1 \equiv & d \omega^1 + \omega^2 \wedge \omega^3 = i\, (u \wedge v - \bar{u} \wedge \bar{v}), \\[2mm]
	\Omega^2 \equiv & d \omega^2 + \omega^3 \wedge \omega^1 = u \wedge v + \bar{u} \wedge \bar{v}, \\[2mm]
	\Omega^3 \equiv & d \omega^3 + \omega^1 \wedge \omega^2 = i\, (u \wedge \bar{u}+ v \wedge \bar{v}). 
\end{align}
We can then recognize that the Killing vector of the gauged isometry is $\vec{k}_{\Lambda} = e_{\Lambda} \partial_w$, with prepotential
\begin{equation}
	P_{\Lambda} = \left(\begin{array}{c}
	0 \\ 0 \\ - e_{\Lambda} \rho
	\end{array}\right),
\end{equation}
so that 
\begin{equation}
	V = \left(U^{\Lambda \Sigma} - 3 L^{\Lambda} \bar{L}^\Sigma\right)P_{\Lambda}^x P_{\Sigma}^x + 4 L^{\Lambda} \bar{L}^\Sigma k_{\Lambda}^u h_{uv} k_{\Sigma}^v = \frac12 \, (\Im F)^{\Lambda \Sigma} e_{\Lambda} e_{\Sigma} \, \rho^2 - 2 \, \rho |L^\Lambda e_{\Lambda}|^2,
\end{equation}
where we also used (\ref{Udef}) and identified
\begin{equation}
	\rho = \chi_T/2.
\end{equation}

This match proves that the off-shell gauging of the tensor multiplet by means of the couplings in (\ref{acttens}) produces the gauging of a U(1) isometry of the hypermultiplet scalar manifold, which would otherwise be possible only on-shell.
It would be interesting to see if it is possible to extend this construction to more general gaugings and to different isometries of the same or of other hyper scalar manifold.
This would however require the possibility to describe different scalars that transform linearly under the action of the gauge isometry by means of dual tensor fields.
In turn, this needs a generalization of the construction presented in section \ref{sub:tensor_multiplet_action} to allow for more general tensor potentials $\chi_T$, which at present seems beyond reach.

% section gauging (end)

\bigskip
\section*{Acknowledgments}

\noindent We would like to thank A.~Van Proeyen and especially G.~Bossard for discussions and comments on this project.
This work is supported in part by the Padova University Project CPDA144437.

\appendix

\section{Conventions} % (fold)
\label{sec:conventions}

We use the mostly plus convention for the metric, with a totally antisymmetric symbol $\epsilon_{0123} = 1$.

Spinors are Majorana, but we write them in chiral notation like in \cite{Andrianopoli:1996cm}:
\begin{align}
	\gamma_5 \psi^A &= \psi^A, \quad \gamma_5 \phi^A = - \phi^A, \\[2mm]
	\gamma_5 \psi_A &= -\psi_A, \quad \gamma_5 \phi_A =  \phi_A, \\[2mm]
	\gamma^{abcd} &= -i \epsilon^{abcd} \gamma_5.
\end{align}
This also means that charge conjugation acts as
\begin{equation}
	(\psi^A)^c = \psi_A, \quad (R^A{}_B)^c = - R^B{}_A.
\end{equation}
Self and anti-selfdual parts of antisymmetric tensors are extracted by
\begin{equation}
	t^{\pm}_{ab} = \frac12\,\left(\delta^{cd}_{ab} \pm \frac{i}{2} \epsilon_{ab}{}^{cd}\right) t_{cd}.
\end{equation}
There are a number of interesting relations on the SU(2) indices, which we used in our calculations:
\begin{align}
	&L^{AB} = \frac{i}{2}\, L^r\, (\sigma^r)^A{}_C \epsilon^{CB}, \qquad
	L_{AB} = \frac{i}{2}\, L^r\,  \epsilon_{AC} (\sigma^r)^C{}_B, \\[2mm]
	&X^{AC} V_{CB} = X^{AC} V_{BC} - \delta^A_B X^{CD} V_{CD} \quad \rm{if} \quad X^{AC} = - X^{CA},\\[2mm]
	&X_{[AB]} = \frac12 \epsilon_{AB} \, \epsilon^{CD} X_{CD}, \\[2mm]
	&{\cal V}^C \Phi^B_i = {\cal V}^B \Phi^C_i - {\cal V}^D \Phi^E_i \epsilon_{DE} \epsilon^{BC}.
\end{align}
	
Also, some relations between fermions are also useful to check the closure of the Bianchi identities for the non-linear multiplet:
\begin{align}
	\bar{\psi}^A \gamma^{ab} \psi^C \bar{\tau}_D \gamma_{ab} \tau_B & = -4 \bar{\psi}^C\tau_D \bar{\psi}^A\tau_B -4 \bar{\psi}^A \tau_D \bar{\psi}^C \tau_B, \\[2mm]
	\bar{\psi}^A \gamma^{a} \psi_B \bar{\tau}^C \gamma_{a} \tau_D & = 2 \bar{\psi}_B \tau^C \, \bar{\psi}^A \tau_D.
\end{align}

Finally, we list here some useful Fierz rearrangements and symmetry properties of 1-form spinor bilinears:
\begin{align}
	\bar{\psi}_A \psi_B &= - \bar{\psi}_B \psi_A, \\[2mm]
	\bar{\psi}_A \gamma^a \psi^B &=  \bar{\psi}^B \gamma^a \psi_A, \\[2mm]
	\bar{\psi}_A \gamma^{ab}\psi_B &=  \bar{\psi}_B \gamma^{ab} \psi_A, \\[2mm]
	\psi_A \bar{\psi}_B &= \frac12\, \bar{\psi}_B \psi_A\, - \frac18\, \bar{\psi}_B \gamma_{ab} \psi_A \, \gamma^{ab}, \\[2mm]
	\psi_A\, \bar{\psi}^B &= \frac12\, \bar{\psi}^B \gamma^a \psi_A \, \gamma_a.
\end{align}

% section conventions (end)

%%%%%%%%%%%%%%%%%%%%%%%%%%%%%%%%%%%%%%%%%%%%%%%%%%%%%%%%%%%%%%

\end{document}